\documentclass[preprint,prd,aps,floatfix,preprintnumbers,superscriptaddress,bibnotes,nofootinbib]{revtex4-1}

\usepackage{epsfig}
\usepackage{amsmath}
\usepackage{latexsym}
\usepackage[psamsfonts]{amssymb}
\usepackage{graphicx}
\usepackage{ulem}
\usepackage{longtable}
\usepackage{epstopdf}
\usepackage{bm}% bold math
\usepackage {tablefootnote} % footnote for Table
\usepackage{color}

\usepackage{algpseudocode}   % pseudocode
\usepackage{algorithmicx}

\usepackage{algorithm}
\usepackage{algpseudocode}

\makeatletter
\newenvironment{BoxedAlgorithm}[1][H]
{%
  \begin{algorithm}[#1]
  \begin{center}
  \begin{minipage}{0.95\linewidth}
  \hrule
  \vspace{2mm}
}
{%
  \vspace{2mm}
  \hrule
  \end{minipage}
  \end{center}
  \end{algorithm}
}
\makeatother

\usepackage{comment}

\newcommand{\be}{\begin{equation}}
\newcommand{\ee}{\end{equation}}
\newcommand{\bea}{\begin{eqnarray}}
\newcommand{\eea}{\end{eqnarray}}
\newcommand{\bi}{\begin{itemize}}
\newcommand{\ei}{\end{itemize}}

\begin{document}
\preprint{KEK-TH-2772}
%-- title ---

\title{
Unbiased Krylov subspace method for the extraction of ground state from lattice correlators
}

%-- author list ---
%
\author{Ryutaro Tsuji}
\email[E-mail: ]{rtsuji@post.kek.jp}
\affiliation{High Energy Accelerator Research Organization (KEK), Ibaraki 305-0801, Japan}
\author{Shoji Hashimoto}
\affiliation{High Energy Accelerator Research Organization (KEK), Ibaraki 305-0801, Japan}
\author{Ryan Kellermann}
\affiliation{High Energy Accelerator Research Organization (KEK), Ibaraki 305-0801, Japan}

\date{\today}
%%%%%%%%%%%%% ABSTRACT %%%%%%%%%%%%%%%%%%%%%%
%-- abstract ---
\begin{abstract}

Ground-state energy and matrix element are reconstructed from correlators in lattice QCD by diagonalizing transfer matrix $\hat{\mathcal{T}}$ within the Krylov subspace spanned by $\hat{\mathcal{T}}^n|\chi\rangle$, where $|\chi\rangle$ is a state generated by an interpolating field on the lattice. In numerical applications, this strategy is spoiled by statistical noise. To circumvent the problem, we introduce a low-rank approximation based on a singular-value decomposition of a matrix made of the correlators. The associated bias is eliminated by an extrapolation to the limit of vanishing variance of energy eigenvalue. The strategy is tested using a set of mock data as well as real data of $K$ and $D_s$ meson correlators.

\end{abstract}

%%%%%%%%%%%%%%%%%%%%%%%%%%%%%%%%%%%%%%%%%%
\maketitle

%--- main text -------------------------------------------------------
 
%%%%%%%%%%%%%%  SEC 1  %%%%%%%%%%%%%%%%%%%%%%%%
\section{Introduction}
\label{sec:introduction}

In lattice Quantum Chromodynamics (QCD) calculations, the properties of hadrons are obtained from correlation functions of local operators. Lowest-lying hadronic states are extracted from the correlator $C(t)$ at large imaginary time separation $t$ using exponential time evolution $e^{-E_it}$ of a state of energy $E_i$. The excited states with energy $E_i$ ($E_i>E_0$) decay rapidly and only the lowest-lying state with $E_0$ survives for sufficiently large $t$. This poses a problem because the statistical noise in the Monte Carlo simulation typically grows exponentially for large time separations \cite{Parisi:1983ae,Lepage:1989hd}. The signal must be extracted before it is masked by the noise, but after the excited states are sufficiently suppressed.
This problem is most significant for baryons and heavy hadrons for which the signal-to-noise ratio declines rapidly. An example is the computation of the nucleon form factor, {\it e.g.} as studied in \cite{Aoki:2025taf}, where the saturation of the ground state is realized only slowly, or the computation of the Compton amplitude, such as those required in the inclusive decays, see \cite{Gambino:2020crt} for example, where two current insertions are needed between the ground states.

One possible strategy to tackle this issue is to design operators that couple more strongly to the ground state and less to the excited states. The variational method \cite{Michael:1985ne,Luscher:1990ck} is proposed to optimize the operator, which is formulated using the Generalized Eigenvalue Problem (GEVP). The re-based GEVP~\cite{RBC:2023xqv} reduces the set of bases and further optimizes the method.
However, it does not assure that unwanted excited states are sufficiently eliminated with a given set of bases.

Another approach is to determine the eigenvalues and eigenvectors of the transfer matrix that describe the time evolution of the lattice correlators \cite{Wagman:2024rid,Hackett:2024xnx,Hackett:2024nbe,Chakraborty:2024scw, Ostmeyer:2024qgu, Abbott:2025yhm, Ostmeyer:2025igc}. Although the explicit form of the transfer matrix $\mathcal{\hat{T}}$ is not known, the correlators can be viewed as matrix elements of $\mathcal{\hat{T}}^t$ ($t$ is a positive integer), that span a Krylov subspace.
The Lanczos algorithm or the GEVP method is applied to obtain the eigenvalues and eigenvectors corresponding to individual eigenstates. The method itself is valid as far as large enough time separation $t$'s are included, but a problem arises when the data have statistical noise. It appears as spurious eigenvalues that do not correspond to any states, and the authors of \cite{Wagman:2024rid,Hackett:2024nbe,Hackett:2024xnx,Abbott:2025yhm} introduced a generalization of the Cullum-Willoughby test \cite{CULLUM1981329,Cullum:2002} with the Lanczos algorithm to eliminate them.

The Prony method and its extensions 
were also developed and applied to evaluate energy  levels \cite{10.1007/3-540-28504-0_14,Beane:2009kya,Cushman:2019tcv,PhysRevE.102.043303,Fleming:2023zml}. Indeed, the Prony method has been shown to have an algebraic relation to the GEVP method constructed using the Hankel matrix $H_{ij}(t)=C(i+j+t)$ of the lattice correlator $C(t)$, and it is  equivalent to the Lanczos alrorithm \cite{Fischer:2020bgv,Ostmeyer:2024qgu,Chakraborty:2024scw}. In order to circumvent spurious eigenvalues, various approaches are proposed in the context of the GEVP formulation, {\it e.g.} a noise filtering method assuming the Gaussian distribution of eigenvalues \cite{Chakraborty:2024scw}. In a recent publication, a truncation of the Hankel matrix is shown to provide a very close approximation of the $\chi^2$-function \cite{Ostmeyer:2025igc}.

Although no fitting of lattice correlators is necessary to estimate $E_i$'s in 
the approaches to diagonalize $\hat{\mathcal{T}}$
and thus no {\it fake plateau} problem arises due to bad choices of fitting range, the spurious eigenvalues may introduce a different kind of bias.
Associated uncertainties are estimated by a (nested) bootstrap resampling method in \cite{Wagman:2024rid,Hackett:2024xnx,Hackett:2024nbe,Chakraborty:2024scw,Ostmeyer:2024qgu, Abbott:2025yhm, Ostmeyer:2025igc}.
The works that utilized the Lanczos algorithm \cite{Wagman:2024rid,Hackett:2024nbe,Hackett:2024xnx} estimated the error as a bootstrap variance after removing outliers. Namely, the
bootstrap samples are eliminated when the number of survived eigenstates after the Cullum-Willoughby test is lower than an imposed requirement. In addition, rigorous two-sided bounds on $E_i$ associated with numerical Lanczos convergence properties are found.
The error in the low-rank approximation approach \cite{Ostmeyer:2025igc} is also estimated in a similar manner, while no bounds are introduced unlike the Lanczos-based approach.
On the other hand, the filtering method \cite{Chakraborty:2024scw} adopts the 68\% percentile of the full samples to estimate the error without removing any outliers, because the filtering is imperfect in practice and the distribution of $E_i$ is distorted such that the standard deviation assuming the normal distribution would not give a robust estimate. 

In this paper, we propose a strategy for identifying energy eigenvalues and eigenvectors using a low-rank approximation to avoid spurious eigenmodes.
Unlike the previous works, a potential bias due to the low-rank approximation is corrected using the data themselves.
The results for the energy, for example, follow the normal distribution.
We call this the unbiased Krylov subspace method. The state vectors are chosen on the basis of the singular value decomposition (SVD). Mean values and statistical uncertainties are estimated using the bootstrap resampling method with respect to the number of configurations (see Appendix~\ref{app:bootstrap_resampling_to_evaluate_statistical_uncertainty} for details). A potential systematic bias due to the truncation of the Krylov subspace is eliminated by an extrapolation in the ``variance'' of the eigenvalue \cite{doi:10.1143/JPSJ.69.2723,doi:10.1143/JPSJ.70.2287,Wu:2023fgp}, using the fact that the eigenstates of $\mathcal{T}^2$ are not necessarily the same as those of $\mathcal{T}$ within the low-rank approximation. Thus, unbiased estimates of eigenvalues and eigenvectors are realized by extrapolating to the limit that the eigenstates match.

This paper is organized as follows.
In Section~\ref{sec:method}, we first present our strategy for calculating the hadron mass and the matrix element. The Transfer matrix GEVP (TGEVP) approach \cite{Chakraborty:2024scw, Abbott:2025yhm, Ostmeyer:2025igc}, which evaluates the energy levels from the correlation function, is extended to evaluate the matrix element.
Section~\ref{sec:spurious_eigenvalues_and_the_energy-variance_extrapolation} discusses the method of identifying the physical eigenvalues and eigenvectors.
In Section~\ref{sec:numerical_test_with_mock_data}, we show the numerical test using mock data with and without noise.
Subsequently, Section~\ref{sec:application_to_realistic_lattice_data} treats realistic lattice data for $K$ and $D_s$ mesons provided by the JLQCD Collaboration. A summary and perspective are given in Section~\ref{sec:summary}.

In this paper, hereafter, the matrix elements are given in the Euclidean metric convention. $\gamma_5$ is defined by $\gamma_5 \equiv \gamma_1\gamma_2\gamma_3\gamma_4=-\gamma_5^M$, which has the opposite sign relative to that in the Minkowski convention ($\vec{\gamma}^M=i\vec{\gamma}$ and $\gamma_0^M=\gamma_4$).

%%%%%%%%%%%%%%  SEC 2  %%%%%%%%%%%%%%%%%%%%%%%%
\section{Diagonalization of the transfer matrix}
\label{sec:method}

In this section, we outline the strategy to diagonalize the transfer matrix to obtain the energy eigenvalues from the lattice correlators. It has already been covered by \cite{Abbott:2025yhm}, for example, but here we iterate to define our notation.

Let us consider a zero-momentum projected two-point correlation function
\begin{align}
    \label{eq:twopt}
    C(t)
    & =
    \sum_{\bm{x}}
    \langle \Omega |
    O_{\psi}(t,\bm{x})
    \overline{O}_\chi(0,\bm{0})
    | \Omega \rangle
    \nonumber\\
    & =
    \langle \psi |  
    \hat{\mathcal{T}}^{t}
    | \chi \rangle,
\end{align}
where state $|\chi\rangle = \overline{O}_{\chi}(0,\bm{0})|\Omega\rangle$ is created by an interpolating operator $\overline{O}_{\chi}$. The operator to annihilate, $O_\psi$, may or may not be the same operator as $O_\chi$.
The time evolution is induced by the transfer matrix $\hat{\mathcal{T}} = \mathrm{e}^{-\hat{\mathcal{H}}}$ with the Hamiltonian operator $\hat{\mathcal{H}}$ of the theory.
In lattice QCD calculations, the two-point correlation function $C(t)$ is obtained only at a finite number of discrete time separations $t=0$, $a$, $2a$, $\cdots$, $(N_T-1)a$, for a lattice spacing $a$ and the temporal extent of the lattice $N_T$.
For simplicity, we set the lattice spacing $a$ to be unity, and write the correlation function (\ref{eq:twopt}) as
\begin{align}
    \label{eq:twopt_latticeunit}
    C(n) = \langle \psi | \hat{\mathcal{T}}^n| \chi \rangle.
\end{align}
The explicit form of the transfer matrix, or the Hamiltonian, is not necessary in the following analysis; we only use its matrix elements $C(n)$ obtained through the Monte Carlo estimate of the correlator.

\subsection{Reconstruction of eigenstates within Krylov subspace}
The objective is to extract the largest eigenvalues of $\hat{\cal T}$ or the lowest eigenvalues of $\hat{\cal H}$, which correspond to the low-lying hadronic energy created by $\overline{O}_{\chi}$ and annihilated by $O_\psi$. We construct them from the finite set of $C(n)$, which amounts to diagonalizing the matrix $\hat{\cal T}$ in the available subspace. Namely, we consider the Krylov subspace given by
\begin{align}
    \label{eq:krylov_subspace}
    \mathcal{K}_m^{\chi}
    =\mathrm{Span}\{
    |\chi\rangle,
    \hat{\cal T}|\chi\rangle,
    \cdots,
    \hat{\cal T}^m|\chi\rangle
    \},
\end{align}
and
\begin{align}
    \label{eq:psikrylov_subspace}
    \mathcal{K}_m^\psi
    =\mathrm{Span}\{
    |\psi\rangle,
    \hat{\cal T}|\psi\rangle,
    \cdots,
    \hat{\cal T}^m|\psi\rangle
    \},
\end{align}
with the dimension of the subspace $m$ satisfying $2m+2\le N_T$ to avoid a wrap-around effect over the time extent of the lattice.
The eigenvalue problem of the transfer matrix $\hat{\mathcal{T}}$ is represented as
\begin{align}
    \label{eq:eigenvalue_problem_original}
    \hat{\mathcal{T}} | x_n \rangle
    =
    \lambda_n | x_n \rangle,
\end{align}
where $\lambda_n$ and $|x_n\rangle$ are the $n$-th eigenvalue and the corresponding right-eigenvector, respectively.
The eigenvalue problem of $\hat{\mathcal{T}}$ can then be approximately expressed as a generalized eigenvalue problem (GEVP) within $\mathcal{K}_m^{\chi,\psi}$ \cite{Chakraborty:2024scw}, which is called the Transfer matrix GEVP (TGEVP) approach.

We denote an approximation of $|x_n\rangle$ in the Krylov subspace $\mathcal{K}_m^\chi$, as $|x_n^{(m),R}\rangle$, and write as a linear combination of $\hat{\cal T}^j|\chi\rangle$'s as
\begin{align}
    \label{eq:righteigenvector}
    |x_n^{(m),R}\rangle
    =
    \sum_{j=0}^{m}(x_n^{(m),R})_j \hat{\mathcal{T}}^j| \chi \rangle,
\end{align}
with coefficients $(x_n^{(m),R})_j$.
Similarly, the left-eigenvector in the $\mathcal{K}_m^\psi$ is represented as 
\begin{align}
    \label{eq:lefteigenvector}
    \langle x_n^{(m),L} |
    =
    \sum_{j=0}^{m}(x_n^{(m),L})_j \langle \psi |\hat{\mathcal{T}}^j.
\end{align}
Sandwiching $\hat{\cal T}$ by $\langle \psi | \hat{\mathcal{T}}^i$ and the right-eigenvector (\ref{eq:righteigenvector}), one obtains a GEVP 
\begin{align}
    \label{eq:gevp_right}
    \sum^{m}_{j=0}
    C^{+(m)}_{ij}(x^{(m),R}_n)_j
    =
    \lambda_n^{(m)}
    \sum^{m}_{j=0}
    C^{(m)}_{ij}(x^{(m),R}_n)_j,
\end{align}
where 
\begin{align}
    \label{eq:matrix_C+}
    C^{+(m)}_{ij}
    & =
    \langle \psi | \hat{T}^i \hat{\mathcal{T}} \hat{\mathcal{T}}^j | \chi \rangle
    =
    C(i+j+1),
    \\
    \label{eq:matrix_C}
    C^{(m)}_{ij}
    & =
    \langle \psi | \hat{\mathcal{T}}^i \hat{\mathcal{T}}^j | \chi \rangle
    =
    C(i+j),
\end{align}
for $0\le i,j \le m$.
In a similar manner, the GEVP for the left-eigenvector is obtained as
\begin{align}
    \label{eq:gevp_left}
    \sum^{m}_{j=0}
    (x^{(m),L}_n)_j C^{+(m)}_{ji}
    =
    \lambda_n^{(m)}
    \sum^{m}_{j=0}
    (x^{(m),L}_n)_j C^{(m)}_{ji}.
\end{align}
When the dimension of the Krylov subspace is large enough, the resulting eigenvalue $\lambda_n^{(m)}$ should approximate the exact eigenvalue $\lambda_n$ well, and the energy eigenvalue is approximated by
\begin{align}
    \label{eq:eigenenergy}
    E^{(m)}_n = -\mathrm{ln}\lambda_n^{(m)}.
\end{align}
The eigenstate of the transfer matrix $\mathcal{T}$ can also be constructed from the correlators:
\begin{align}
    \label{eq:energy_eigenstate_vector}
    | E_n^{(m),R} \rangle
    & =
    \frac{|x_n^{(m),R}\rangle}{\sqrt{\langle x^{(m),L}_n|x^{(m),R}_n\rangle}}
    \nonumber
    \\
    &
    =
    \frac{
    \sum_{j=0}^{m}
    (x_i^{(m),R})_j
    \hat{\mathcal{T}}^j|\chi\rangle
    }{
    \sqrt{
    \left[
    \sum^m_{i,j=0}
    (x_n^{(m),L})_i (x_n^{(m),R})_j C(i+j)
    \right]
    }
    },
    \\
    \langle E_n^{(m),L} |
    & =
    \frac{\langle x_n^{(m),L}|}{\sqrt{\langle x^{(m),L}_n|x^{(m),R}_n\rangle}} 
    \nonumber
    \\
    & 
    =
    \frac{
    \sum_{j=0}^{m}
    (x_i^{(m),L})_j
    \langle \psi|
    \hat{\mathcal{T}}^j
    }{
    \sqrt{
    \left[
    \sum^m_{i,j=0}
    (x_n^{(m),L})_i (x_n^{(m),R})_j C(i+j)
    \right]
    }
    }.
\end{align}
These eigenstate vectors are normalized as $\langle E^{(m),L}_n | E^{(m),R}_n \rangle = 1$. (Note the difference from the canonical normalization for which the right-hand side becomes $2E_n$.)
For a symmetric correlator, {\it i.e.} $\psi=\chi$, the left eigenvector is a complex  conjugate of the right eigenvector.
These eigenstates can be used to reconstruct the correlation function of a specific state as discussed in the next subsection and to evaluate the hadron matrix elements.

\subsection{Hadron matrix elements with GEVP}
\label{ssec:hadron_matrix_elements_with_GEVP}
We consider the hadron matrix elements of an operator $\hat{\mathcal{J}}$, {\it i.e.}
$
    J_{fi}=\langle E_f | \hat{\mathcal{J}} | E_i \rangle
$,
for the states $|E_i\rangle$ and $\langle E_f|$.
They are extracted from a simultaneous analysis of two- and three-point correlation functions. For zero-momentum projected states, we can write the three-point correlation function as
\begin{align}
    \label{eq:threept_latticeunit}
    C^{\mathrm{3pt}}(\sigma, \tau) 
    & =
    \langle \psi | \hat{\mathcal{T}}^\sigma \hat{\mathcal{J}} \hat{\mathcal{T}}^\tau | \chi \rangle
    =
    \sum_{f,i}Z^{*}_f Z_i J_{fi} \mathrm{e}^{-E_f \sigma - E_i\tau},
\end{align}
where $Z_i=\langle E_i|\chi\rangle$ and $Z_f=\langle E_f|\psi\rangle$.
An extension to the case of non-zero momentum states is straightforward.
Through GEVP of the transfer matrix within the Krylov subspace, (\ref{eq:gevp_right}) and (\ref{eq:gevp_left}), the eigenvectors $|E_n^{(m),R}\rangle$ and $\langle E_n^{(m),L}|$ can be constructed as discussed in the previous subsection. 
They can be used to construct the matrix elements, and the results are written in terms of an appropriate combination of three-point correlation functions, {\it i.e.}
\begin{align}
    \label{eq:derive_hadron_matrix_element_sameop}
    J^{(m)}_{fi} 
    & =
    \langle E^{(m),L}_f | \hat{\mathcal{J}} | E^{(m),R}_i \rangle
    \nonumber\\
    & =
    \frac{
    \sum_{k,l=0}^{m}
    (x_f^{(m),L})_k 
    \langle \chi | \hat{\mathcal{T}}^{k} \hat{\mathcal{J}} \hat{\mathcal{T}}^{l}| \chi \rangle
    (x_i^{(m),R})_l
    }{\sqrt{\langle x^{(m),L}_f|x^{(m),R}_f\rangle}\sqrt{\langle x^{(m),L}_i|x^{(m),R}_i\rangle}}
    \nonumber\\
    & =
    \frac{
    \sum_{k,l=0}^{m}
    (x_f^{(m),L})_k 
    C^\mathrm{3pt}(k,l)
    (x_i^{(m),R})_l
    }{\sqrt{\langle x^{(m),L}_f|x^{(m),R}_f\rangle}\sqrt{\langle x^{(m),L}_i|x^{(m),R}_i\rangle}}.
\end{align}
In particular, the ground-state matrix element with identical initial and final states is obtained as
\begin{align}
    \label{eq:ground-state_matrix_element}
    J^{(m)}_{00} 
    =
    \frac{
    \sum_{k,l=0}^{m}
    (x_0^{(m),L})_k 
    C^\mathrm{3pt}(k,l)
    (x_0^{(m),R})_l
    }{
   \langle x^{(m),L}_0|x^{(m),R}_0\rangle 
    }.
\end{align}

The form of (\ref{eq:derive_hadron_matrix_element_sameop}) and (\ref{eq:ground-state_matrix_element}) implies that the whole set of three-point functions $C^\mathrm{3pt}(k,l)$ with $0\leq k,l\leq m$ is necessary to reconstruct the matrix elements. This is different from the standard analysis of form factors, that is, one fixes the source-sink separation $t_\mathrm{sep}$ so that only the combinations of $k$ and $l$ satisfying $k+l=t_\mathrm{sep}$ are used. In some cases, the dependence on $t_\mathrm{sep}$ is also studied having several values of $t_\mathrm{sep}$, but the computation required to apply (\ref{eq:derive_hadron_matrix_element_sameop}) and (\ref{eq:ground-state_matrix_element}) is much larger. This is the cost of constructing the eigenstates of the transfer matrix (or Hamiltonian) without relying on the assumption that the lattice correlators are saturated by a few exponential functions in the given time range.

The truncation of the Krylov subspace $\mathcal{K}_m^{\chi,\psi}$ is a potential source of the systematic effect. The truncation is unavoidable not only because of the computational cost, but the available range of time separation is limited by the exponential growth of statistical noise for large time separations \cite{Parisi:1983ae,Lepage:1989hd}. In the next section, we discuss in some detail how the statistical noise affects the analysis and how one can analyze with a limited range of time separations.

%%%%%%%%%%%%%%  SEC 3  %%%%%%%%%%%%%%%%%%%%%%%%
\section{Spurious eigenvalues and the energy-variance extrapolation}
\label{sec:spurious_eigenvalues_and_the_energy-variance_extrapolation}

Although the strategy outlined in the previous section provides a theoretically valid framework to reconstruct energy eigenstates from lattice correlators, there is a limitation owing to the truncation of the Krylov subspace. The Krylov subspace has to be truncated before it is overwhelmed by statistical noise. Without truncation, the role of each eigenvalue (and eigenvector) fluctuates too much, which is sometimes called {\it spurious} eigenvalues, a manifestation of the Bauer--Fike theorem~\cite{Bauer1960NormsAE}.

In the presence of statistical noise, the Krylov subspace $\mathcal{K}_m^{\chi,\psi}$ in which the eigenstates are constructed also fluctuates. In particular, a problematic situation could occur in which the $(j+1)$-th vector $\hat{\cal T}^{j+1}|\chi\rangle$ does not contain a component orthogonal to $\mathcal{K}_j$ with sufficient statistical significance. The resulting eigenvalue is called spurious. 
It may even violate the necessary condition $0\leq\lambda_{j+1}^{(m)}<1$ required for the transfer matrix, or appear as the lowest energy state with some probability, which indicates that the spurious eigenvalues may distort the distribution of the ground-state energy.

In order to eliminate such problems and to make the method more well-defined, we propose an unbiased Krylov subspace method. It consists of two key components: (i) a low-rank approximation with a singular value decomposition to make the statistical fluctuation of the Krylov subspace under control, and (ii) an eigenvalue-variance extrapolation, which eliminates the associated bias due to the low-rank approximation. This method provides a framework to extract the ground-state mass and the hadron matrix element in a systematic manner. The algorithms we have implemented are summarized in Algotithm~\ref{alg:spectroscopy} and \ref{alg:hadron_matrix_element}.

\subsection{Singular value decomposition (SVD)}
\label{ssec:a_singular_value_decomposition}
The TGEVP defined by the Prony formalism, (\ref{eq:gevp_right}) and (\ref{eq:gevp_left}), may be expressed as
\begin{align}
    C^{+(m)}|x_n^{(m)}\rangle =\lambda^{(m)}_n C^{(m)} |x_n^{(m)}\rangle,
\end{align}
with $(m+1)\times(m+1)$ complex matrices $C^{+(m)}$ and $C^{(m)}$, and $|x_n^{(m)}\rangle$ is a $(m+1)$-dimensional complex vector. The SVD of $C^{+(m)}$ transforms the matrix into a product of three matrices: $C^{+(m)}=U^{(m)}\Sigma^{(m)} V^{(m)t}$, where $\Sigma^{(m)}$ is a diagonal matrix with non-negative entries of singular values, 
$\Sigma^{(m)} = \mathrm{diag}(\sigma_0,\sigma_1,\dots,\sigma_r,\sigma_{r+1},\dots,\sigma_{m})$
sorted from high to low ($\sigma_0\ge\sigma_1\ge\dots$).
Keeping only the highest singular values $(r+1)$, $C^{+(m)}$ can be approximated by a $(m+1)\times (m+1)$ matrix of $C^{+(m)}_r= U^{(m)}_r\Sigma^{(m)}_r V^{(m)t}_r$ with $\Sigma^{(m)}_r = \mathrm{diag}(\sigma_0,\sigma_1,\dots,\sigma_r, 0, \cdots)$ such that $C^{+(m)}\simeq C^{+(m)}_r$ up to the ignored components of small singular values.
And, $U^{(m)}_r$ and $V_r^{(m)t}$ are orthogonal matrices to diagonalize $C^{+(m)}_r$; they are identical in our case where $C^{+(m)}$ and $C^{(m)}$ 
constructed from $C(t)$ as (\ref{eq:matrix_C+}) and (\ref{eq:matrix_C})
are real and symmetric.
This is called the low-rank approximation. 
If one applies the SVD to $C^{(m)}$ instead of $C^{+(m)}$, one can obtain a similar formulation.

The SVD, $C^{+(m)}=U^{(m)}\Sigma^{(m)} V^{(m)t}$, depends on statistical samples. We employ the bootstrap method for statistical analysis, and the SVD is performed for each bootstrap sample using the central values of the elements of $C^{+(m)}$. Thus, the resulting singular values, $\sigma_0$, $\sigma_1$, ..., have some distribution over the bootstrap samples. When their statistical fluctuation is too large, the distribution of a singular value $\sigma_j$ ($j<m$) overlaps that of $\sigma_{j+1}$, obscuring even the ordering of singular values. We call such singular values {\it spurious}. The analysis breaks down in such a case because the Krylov subspace can be vastly different among the bootstrap samples. We truncate the Krylov subspace such that this problem does not occur. That is, we set $r$ sufficiently small to eliminate spurious singular values.

In the truncated $(r+1)$-dimensional Krylov subspace, the GEVP is formulated as
\begin{align}
\label{eq:lowrank_gevp_R}
\sum_{t=0}^r
(\Sigma^{(m)}_{r})_{st} (y^{(m),R}_{n})_t
& ={\lambda}^{\prime(m)}_{n}
\sum_{t=0}^r
(U_r^{(m)t} C^{(m)} V^{(m)}_r)_{st} (y^{(m),R}_{n})_t,
\\
\label{eq:lowrank_gevp_L}
\sum_{t=0}^r
 (z^{(m),L}_{n})_t(\Sigma^{(m)}_{r})_{ts}
 & ={\lambda}^{\prime(m)}_{n}
\sum_{t=0}^r
(z^{(m),L}_{n})_t
(U_r^{(m)t} C^{(m)} V^{(m)}_r)_{ts}
\end{align}
with
\begin{align}
\label{eq:lowrank_rightsolution}
(y^{(m),R}_{n})_t & =
\sum_{j=0}^m (V^{(m)t}_r)_{tj}(x^{(m),R}_n)_j,\\
(z^{(m),L}_{n})_t & =
\sum_{j=0}^m (x^{(m),L}_n)_j(U^{(m)}_r)_{jt},
\end{align}
where ${\lambda}^{\prime(m)}_n=\lambda^{(m)}_n$ if $r=m$.
We denote the maximal value of $r$ that we take as $r_\mathrm{max}$.
Hereafter, the eigenvalues and matrix elements with and without prime indicate the low-rank approximated and original ones, respectively.

The eigenvalues $0 < \lambda^{\prime(m)}_0<1$ are sorted as $\lambda^{\prime(m)}_0 > \lambda^{\prime(m)}_1>\dots$ for each $m$. These provide energies of the hadron $E^{\prime(m)}_n = -\mathrm{ln}\lambda_n^{\prime(m)}$ and in particular the ground-state energy corresponds to $E^{\prime(m)}_0 = -\mathrm{ln}\lambda_0^{\prime(m)}$.
The right/left eigenvectors are approximated as
\begin{align}
    |x^{\prime(m),R}_n\rangle
    &
    =
    \sum^{m}_{j=0}
    (x^{\prime(m),R}_n)_j
    \hat{\mathcal{T}}^j|\chi\rangle
    \nonumber
    \\
    \label{eq:lowrank_rightvector}
    & =
    \sum^{m}_{j=0}
    \left[
    \sum_{t=0}^r (V^{(m)}_r)_{jt}(y^{(m),R}_n)_t
    \right]
    \hat{\mathcal{T}}^j|\chi\rangle,\\
    \langle x^{\prime(m),L}_n|
    &
    =
    \sum^{m}_{j=0}
    (x^{\prime(m),L}_n)_j
    \langle \psi |\hat{\mathcal{T}}^j|
    \nonumber
    \\
    \label{eq:lowrank_leftvector}
    & =
    \sum^{m}_{j=0}
    \left[
    \sum_{t=0}^r (z^{(m),L}_n)_t (U^{(m)t}_r)_{tj}
    \right]
    \langle \psi|\hat{\mathcal{T}}^j.
\end{align}
For a symmetric correlator, {\it i.e.} $|\chi\rangle=|\psi\rangle$, the right and left eigenvectors are real and identical, since both $\Sigma^{(m)}_r$ and $U^{(m)t}_r C^{(m)} V^{(m)}_r$ are real and symmetric. We therefore use the notation $|x_n'^{(m)}\rangle\equiv|x_n'^{(m),R}\rangle=|x_n'^{(m),L}\rangle$ in the following.

\subsection{Eigenvalue-variance extrapolation}
\label{ssec:the_eigenvalue-variance_extrapolation}
The low-rank approximation introduces a systematic bias to the results of energy eigenvalues and matrix elements, as it truncates the Hilbert space to represent the states. In order for the method to be useful for practical purposes, a clear prescription is required to estimate and possibly eliminate the associate error. 
In this work, we employ the eigenvalue-variance extrapolation \cite{doi:10.1143/JPSJ.69.2723, PhysRevB.64.024512, doi:10.1143/JPSJ.70.2287,Wu:2023fgp} to estimate the exact eigenvalues from the solutions of (\ref{eq:lowrank_gevp_R}) and (\ref{eq:lowrank_gevp_L}) and the right/left eigenvectors of (\ref{eq:lowrank_rightvector}) and (\ref{eq:lowrank_leftvector}) for several $r$'s.

For the highest eigenvalue $\lambda_0$ of the transfer matrix, we consider the difference between the truncated and untruncated evaluations: 
\begin{align}
\label{eq:eigenvalue_difference}
\delta\lambda_0^{(m)}&\equiv\langle x_0'^{(m)}|\hat{\mathcal{T}}|x_0'^{(m)}\rangle-\langle x_0^{(m)}|\hat{\mathcal{T}}|x_0^{(m)}\rangle\nonumber\\ & = {\lambda}^{\prime(m)}_0 - \lambda_0^{(m)},
\end{align}
where $|x_0^{(m)}\rangle$ is the full eigenvector as constructed in (\ref{eq:righteigenvector}) while that with a prime $|x_0'^{(m)}\rangle$ is truncated at $r$ as in (\ref{eq:lowrank_rightvector}). The suffix to indicate $r$ is suppressed for simplicity. The size $m$ of the Krylov subspace $\mathcal{K}_m$ is fixed throughout the discussion, and the extrapolation of $r\to m$ is needed.
The dependence of $\delta\lambda_0^{(m)}$ on $r$ is, however, unknown a priori, and we take another measure to control the extrapolation as discussed below.

\begin{comment}
We need the extrapolation procedure of $r\to m$ to correct the bias due to the low-rank approximation. However, in practice, it is difficult to formulate the relation between the $r$ and $\delta \lambda_0^{(m)}$. Moreover, it is challenging to define sufficiently large size of $r$ in practice. This is because the distribution of the eigenvalues or the physical quantities constructed by the eigenvectors is distorted by the spurious eigenvalues and corresponding eigenvectors as $r\to m$. Therefore, we introduce another extrapolation procedure to achieve $r\to m$ without using $r$ as an argument.
\end{comment}

The difference $\delta\lambda_0^{(m)}$ (\ref{eq:eigenvalue_difference}) emerges from the truncation of the Krylov subspace and vanishes in the limit of $r\to m$. 
Since $|x_0'^{(m)}\rangle$ is truncated at $r$, it is no longer an exact eigenstate of the transfer matrix; hence, the ``variance'' 
\begin{align}
\label{eq:eigenvalue_variance}
\Delta\lambda^{(m)}_0 \equiv \langle x_0'^{(m)}|\hat{\mathcal{T}}^2|x_0'^{(m)}\rangle-\langle x_0'^{(m)}|\hat{\mathcal{T}}|x_0'^{(m)}\rangle^2
\end{align}
with
\begin{align}
    \label{eq:detail_variance_0}
    \langle x_0'^{(m)}|\hat{\mathcal{T}}^2|x_0'^{(m)}\rangle
    &=
    \sum^{m}_{i,j=0}(x^{\prime(m)})_i(x^{\prime(m)})_j C(i+j+2),
    \\
    \label{eq:detail_variance_1}
    \langle x_0'^{(m)}|\hat{\mathcal{T}}|x_0'^{(m)}\rangle
    &=
    \sum^{m}_{i,j=0}(x^{\prime(m)})_i(x^{\prime(m)})_j C(i+j+1)
\end{align}would be non-zero.
In the leading order of small $\delta\lambda_0^{(m)}$, we can expect $\delta\lambda\propto\Delta\lambda_0$ \cite{doi:10.1143/JPSJ.70.2287}, suggesting that the exact eigenvalue $\lambda_0^{(m)}$ (within $\mathcal{K}_m$) can be obtained by taking the limit $\Delta\lambda^{(m)}_0\to 0$ as $r$ increases.

The relation between $\delta\lambda_0^{(m)}$ and $\Delta\lambda_0^{(m)}$ is evaluated as follows \cite{doi:10.1143/JPSJ.69.2723, doi:10.1143/JPSJ.70.2287}.
The ground state $|x^{\prime(m)}_0\rangle$ in the truncated (or low-rank approximated) subspace can be written as a linear combination of the $(m+1)$ eigenstates of $\mathcal{H}$ in the $(m+1)$-dimensional untruncated subspace. Due to the truncation, $|x^{\prime(m)}_0\rangle$ deviates from the ground state $|x^{(m)}_0\rangle$, contaminated by excited-state contributions. In general, the amount and direction of the deviation in the subspace depend on the truncation. Here, we assume that $|x^{\prime(m)}_0\rangle$ provides a good approximation of $|x^{(m)}_0\rangle$ and that the deviation is along a certain direction $|\bar{x}^{(m)}_0\rangle$, given that the deviation itself is small. We write the ground state in the truncated subspace as
\begin{align}
|x^{\prime(m)}_0\rangle &
= c |x^{(m)}_0\rangle + d |\bar{x}^{(m)}_0\rangle,
\end{align}
where $c^2+d^2=1$ with small $d$.
Parametrizing the matrix elements of the complement as
\begin{align}
    D_1 & \equiv
    \frac{
    \langle \bar{x}^{(m)}_0|
    \mathcal{\hat{T}}
    |\bar{x}^{(m)}_0\rangle
    -
    \langle x^{(m)}_0|
    \mathcal{\hat{T}}
    |x^{(m)}_0\rangle
    }{
    \langle x^{(m)}_0|
    \mathcal{\hat{T}}
    |x^{(m)}_0\rangle
    },\\
    D_2 & \equiv
    \frac{
    \langle \bar{x}^{(m)}_0|
    \mathcal{\hat{T}}^2
    |\bar{x}^{(m)}_0\rangle
    -
    \langle x^{(m)}_0|
    \mathcal{\hat{T}}
    |x^{(m)}_0\rangle^2
    }{
    \langle x^{(m)}_0|
    \mathcal{\hat{T}}
    |x^{(m)}_0\rangle^2
    },
\end{align}
we obtain
\begin{align}
\langle x_0^{\prime(m)}|\hat{\mathcal{T}}|x_0^{\prime(m)}\rangle
& =
\langle x_0^{(m)}|\hat{\mathcal{T}}|x_0^{(m)}\rangle+
d^2 D_1 \langle x_0^{(m)}|\hat{\mathcal{T}}|x_0^{(m)}\rangle,\\
\langle
x_0^{\prime(m)}|\hat{\mathcal{T}}^2|x_0^{\prime(m)}\rangle
& =
\langle x_0^{(m)}|\hat{\mathcal{T}}|x_0^{(m)}\rangle^2+
d^2 D_2 \langle x_0^{(m)}|\hat{\mathcal{T}}|x_0^{(m)}\rangle^2.
\end{align}
Here we use $\langle x_0^{(m)}|\hat{\mathcal{T}}^2|x_0^{(m)}\rangle= \langle x_0^{(m)}|\hat{\mathcal{T}}|x_0^{(m)}\rangle^2$.
Eventually, the $d^2$-dependence of the $\delta \lambda^{(m)}_0$ and $\Delta \lambda^{(m)}_0$ is described as
\begin{align}
    \delta \lambda^{(m)}_0
    & =
    d^2 \lambda^{(m)}_0D_1,\\
    \Delta \lambda^{(m)}_0
    & =
    d^2 (D_2-2D_1) (\lambda^{(m)}_0)^2+ O(d^4).
\end{align}

As long as $|x^{\prime(m)}_0\rangle$ provides a good description of the ground state, the coefficient $d$ is small. Therefore, $\delta\lambda^{(m)}_0\propto\Delta\lambda^{(m)}_0$ up to $O(d^2)$, and an extrapolation to $r\to m$ can be carried out by monitoring $\Delta\lambda^{(m)}_0$ as demonstrated in the following sections.
Finally, the eigenvalue of the ground state $\lambda_0$ for the entire Hilbert space can be estimated by $\lambda_0^{(m)}$ with sufficiently large $m$ where no further $m$ dependence is observed.

Here, it is important to note the behavior of $D_1$ and $D_2$ with respect to the parameters $r$ and $m$, while the $D_1$ and $D_2$ cannot be computed directly for the lattice data. First, the $D_1$ and $D_2$ depend on $m$. This indicates that the slope may depend on $m$, hence the extrapolation is performed for each $m$. 
The $D_1$ and $D_2$ also have $r$-dependence in general, however 
the bias for $\delta\lambda^{(m)}_0$ and $\Delta\lambda^{(m)}_0$ is controlled only by $d^2$ at the leading order, under our assumption. Based on these, a linear extrapolation to $r\to m$ can be carried out as long as $d$ is small.

The similar argument can be applied for the matrix elements. In analogy to (\ref{eq:eigenvalue_difference}), we define the difference $\delta J_{00}^{(m)} = J_{00}^{\prime(m)} - J_{00}^{(m)}$,
where $J_{00}^{\prime(m)}$ is the ground-state matrix element evaluated with the truncated states (\ref{eq:lowrank_rightvector}) and (\ref{eq:lowrank_leftvector})
as
\begin{align}
    \label{eq:lowrank_ground-state_matrix_element}
    J^{\prime(m)}_{00} 
    =
    \frac{
    \sum_{k,l=0}^{m}
    (x_0^{\prime(m),L})_k 
    C^\mathrm{3pt}(k,l)
    (x_0^{\prime(m),R})_l
    }{
   \langle x^{\prime(m),L}_0|x^{\prime(m),R}_0\rangle
    }.
\end{align}
Unlike the discussion on the extrapolation of the eigenvalue, in which $\delta\lambda_0^{(m)}\propto\Delta\lambda^{(m)}_n$, it can be shown that
\begin{align}
    \label{eq:J00}
    \delta J^{(m)}_{00} &
    =
    d^2 J^{(m)}_{00} D_J +
    2cd \langle \bar{x}^{(m)}_0 | \mathcal{\hat{J}}
    | x^{(m)}_0\rangle
\end{align}
with
\begin{align}
    D_J & 
    \equiv
    \frac{
    \langle \bar{x}^{(m)}_0|
    \mathcal{\hat{J}}
    |\bar{x}^{(m)}_0\rangle
    -
    \langle x^{(m)}_0|
    \mathcal{\hat{J}}
    |x^{(m)}_0\rangle
    }{
    \langle x^{(m)}_0|
    \mathcal{\hat{J}}
    |x^{(m)}_0\rangle
    }.
\end{align}
The second term in (\ref{eq:J00}) implies that the leading contribution for small $d$ is linear in $d$ and not quadratic as in $\delta\lambda_0^{(m)}$.
In \cite{doi:10.1143/JPSJ.69.2723,doi:10.1143/JPSJ.70.2287}, it is argued that the relevant matrix element $\langle\bar{x}^{(m)}_0|\mathcal{\hat{J}}|x^{(m)}_0\rangle$ is small so that the leading term of the extrapolation is effectively $\Delta \lambda^{(m)}_0$.
This is the case when the off-diagonal matrix element between the ground state $|x^{(m)}_0\rangle$ and the orthogonal state $|\bar{x}^{(m)}_0\rangle$,
which consists of the excited states, is suppressed for some reason.
In this work, we also investigate the relation between the size of the off-diagonal matrix elements and $\Delta \lambda^{(m)}_0$-dependence for the ground-state matrix element.

\begin{BoxedAlgorithm}   % H = do not split / stay in place
\caption{Spectroscopy}
\label{alg:spectroscopy}
\begin{algorithmic}
\Statex \textbf{Parameters:} dimension of the Krylov subspace $m$
\Statex \textbf{Input:} Bootstrapped correlator $C(t)$ for $t=0, ..., 2m+2$
\Statex \textbf{Output:} Estimated mean and statistical error of $E_0^{(m)}$
\ForAll{Bootstrap samples} 
\State
$C^{+(m)} \gets
	\begin{pmatrix}
		C(1) & C(2) & \cdots & C(m+1) \\
		C(2) & C(3) & \cdots & C(m+2)  \\
		\vdots & \vdots & \ddots & \vdots\\
		C(m+1) & C(m+2) & \cdots & C(2m+1) \\
	\end{pmatrix}$ 
\State $U^{(m)}\Sigma^{(m)} V^{(m)t} \gets C^{+(m)}$ \Comment{\texttt{Singular value decomposition}}
\State $(\sigma_0, \sigma_1, \dots, \sigma_m)\gets \Sigma^{(m)}$
\EndFor
\State Construct a distribution of $\sigma_r^{(m)}/\sigma_0^{(m)}$ from $(\sigma_0, \sigma_1, \dots, \sigma_m)$ using all Bootstrap samples
\State $r_\mathrm{max}\gets$ the largest $r$ s.t. the distributions of $\sigma_{r}^{(m)}/\sigma_0^{(m)}$ and $\sigma_{r-1}^{(m)}/\sigma_0^{(m)}$ do not overlap
\ForAll{Bootstrap samples} \Comment{\texttt{Solve GEVP in truncated Krylov subspace}}
\State
$C^{(m)+} \gets
	\begin{pmatrix}
		C(1) & C(2) & \cdots & C(m+1) \\
		C(2) & C(3) & \cdots & C(m+2)  \\
		\vdots & \vdots & \ddots & \vdots\\
		C(m+1) & C(m+2) & \cdots & C(2m+1) \\
	\end{pmatrix}$
\State $C^{(m)} \gets
	\begin{pmatrix}
		C(0) & C(1) & \cdots & C(m) \\
		C(1) & C(2) & \cdots & C(m+1)  \\
		\vdots & \vdots & \ddots & \vdots\\
		C(m) & C(m+1) & \cdots & C(2m) \\
	\end{pmatrix}$ 
\For{$r = 0$ to $r_\mathrm{max}$}
\State $\lambda^{\prime(m)}_0$, $|x^{\prime(m),R}_0\rangle$, $\langle x^{\prime(m),L}_0|$  $\gets$ \textbf{Solve} (\ref{eq:lowrank_gevp_R}) and (\ref{eq:lowrank_gevp_L})
\State $\Delta \lambda^{(m)}_0 \gets |x^{\prime(m),R}_0\rangle$, $\langle x^{\prime(m),L}_0|$, $C(t)$ by (\ref{eq:eigenvalue_variance}), (\ref{eq:detail_variance_0}) and (\ref{eq:detail_variance_1})
\EndFor
\State $\tilde{\lambda}^{(m)}_0 \gets$ \textbf{Extrapolate} $\lambda^{\prime(m)}_0$ as $\Delta \lambda^{(m)}_0 \to 0$ \Comment{\texttt{Bias correction by the extrapolation}}
\State $E_0^{(m)}\gets -\mathrm{ln}\tilde{\lambda}^{(m)}_0$
\EndFor
\State $E[E^{(m)}_0], \mathrm{Var}[E^{(m)}_0]$ from Bootstrap distribution
\end{algorithmic}
\end{BoxedAlgorithm}

\begin{BoxedAlgorithm}   % H = do not split / stay in place
\caption{Hadron matrix element}
\label{alg:hadron_matrix_element}

\begin{algorithmic}[1]
\Statex \textbf{Parameters:} dimension of the Krylov subspace $m$
\Statex \textbf{Input:} Bootstraped correlator $C^{3\mathrm{pt}}(\sigma,\tau)$ for $0\le \sigma, \tau \le 2m+2$
\Statex \hspace{1.4cm}$r_\mathrm{max}, |x^{\prime(m),R}_0\rangle, \langle x^{\prime(m),L}_0|, \Delta \lambda^{(m)}_0$ \Comment{\texttt{From spectroscopy}}
\Statex \textbf{Output:} Estimated mean and statistical error of $J_{00}^{(m)}$
\ForAll{Bootstrap samples} 
\For{$r = 0$ to $r_\mathrm{max}$}
\State $J^{\prime(m)}_{00} \gets |x^{\prime(m),R}_0\rangle, \langle x^{\prime(m),L}_0|, C^{3\mathrm{pt}}(\sigma,\tau)$ by (\ref{eq:lowrank_ground-state_matrix_element})
\EndFor
\State $\tilde{J}^{(m)}_{00} \gets$ \textbf{Extrapolate} $J^{\prime(m)}_{00}$ as $\Delta \lambda^{(m)}_0\to 0$ \Comment{\texttt{Bias correction by the extrapolation}}
\EndFor
\State $E[J^{(m)}_{00}], \mathrm{Var}[J^{(m)}_{00}]$ from Bootstrap distribution
\end{algorithmic}
\end{BoxedAlgorithm}

%%%%%%%%%%%%%%  SEC 4  %%%%%%%%%%%%%%%%%%%%%%%%
\section{Numerical test with mock data}
\label{sec:numerical_test_with_mock_data}

We tested the method described above to isolate the ground state using mock data sets. The two-point correlation function is generated assuming
\begin{align}
    \label{eq:mock2pt}
    C(t) & = \sum_{n=0}^{5}Z_n^2 \mathrm{e}^{-E_nt},
\end{align}
with $E_n=0.1(n+1)$ and $Z_n=1/\sqrt{2E_n}$. Six equally separated states are included with decreasing magnitudes.
The three-point correlation function is set 
\begin{align}
    \label{eq:mock3pt}
    C^{3\mathrm{pt}}(t_\mathrm{sep},t)
    & = \sum_{n,m=0}^{5}Z_mZ_n J_{mn} \mathrm{e}^{-E_m(t_\mathrm{sep}-t)} \mathrm{e}^{-E_nt}
\end{align}
with three types of matrix elements: $J_{mn}^\mathrm{I}=1/(1+m\times n)$, $J_{mn}^\mathrm{II}=1/(1+m\times n)\times0.5$, and $J_{mn}^\mathrm{III}=1/(1+m\times n)\times\delta_{mn}$.
They are introduced to investigate the relation between the size of the off-diagonal matrix elements and the dependence on $\Delta\lambda^{(m)}_0$.

To mimic real analysis, we apply the TGEVP to the normalized two-point correlation function $\bar{C}(t)=C(t+2t_0)/C(2t_0)$, 
with a small but non-zero time separation $t_0$. (We take $t_0=1$ in lattice units.)
This normalization is also taken into account for the evaluation of the hadron matrix element. For simplicity, hereafter, we denote the normalized correlation functions without the bar.

In Section~\ref{ssec:noiseless_mock_data} we analyze the mock data without noise, and then the effect of noise is discussed in Section~\ref{ssec:noisy_mock_data}.

\subsection{Noiseless mock data}
\label{ssec:noiseless_mock_data}
\begin{figure*}[tbp]
\centering
\includegraphics[width=0.7\textwidth,bb=0 30 745 550,clip]{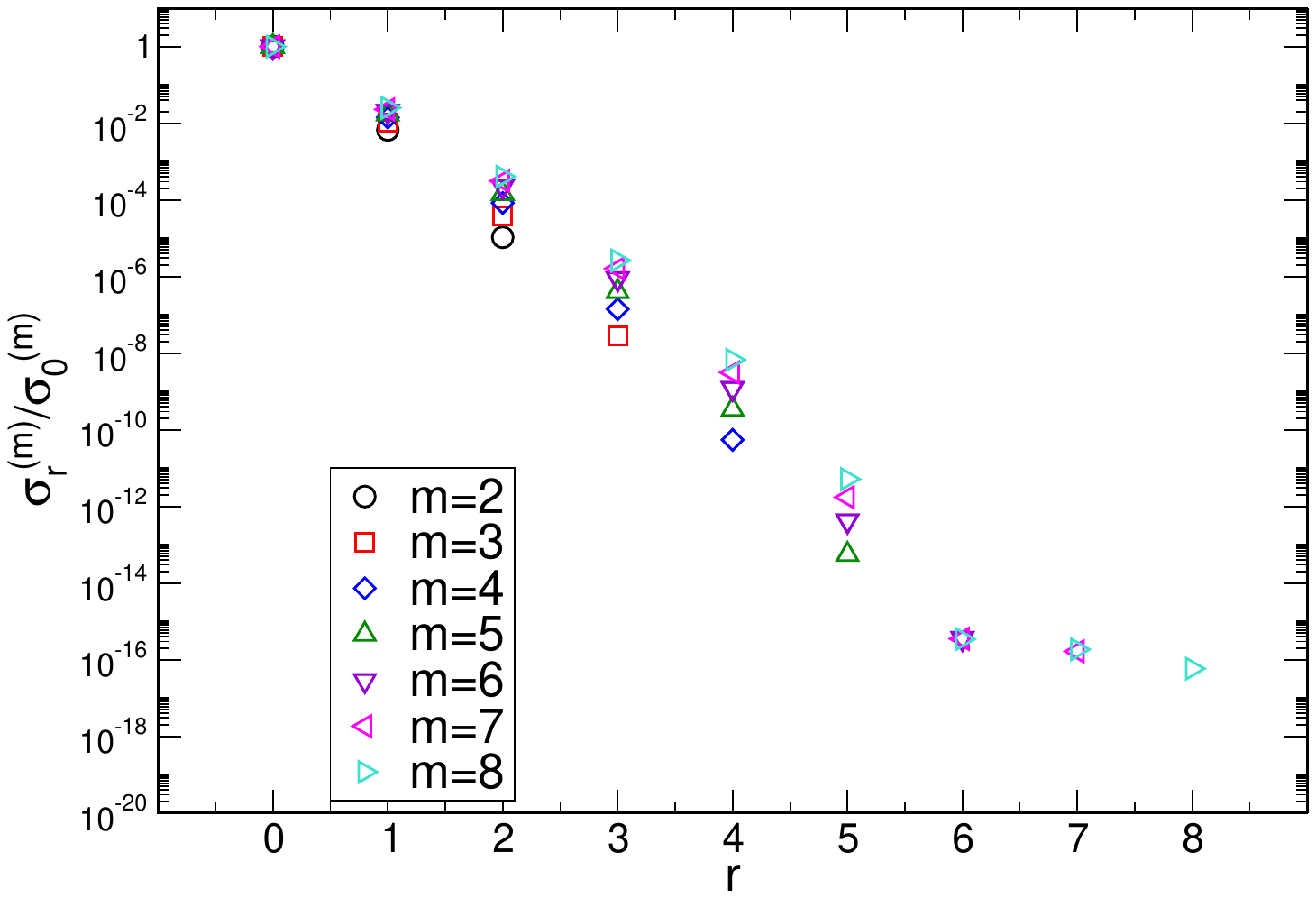}
\caption{
Normalized singular values $\sigma_r^{(m)}/\sigma_0^{(m)}$ of $C^{+(m)}_{ij}$ for the noiseless two-point correlation functions. They are plotted in decreasing order of singular values, for each choice of $m$ denoted in the label.
}
\label{fig:svNoiseless}
\end{figure*}

First, let us examine our strategy with noiseless mock data.
Figure~\ref{fig:svNoiseless} shows the normalized singular values $\sigma_r^{(m)}/\sigma_0^{(m)}$ of $C^{+(m)}_{ij}$ defined in (\ref{eq:matrix_C+}), which is a $(m+1)\times (m+1)$ matrix ($m=2,\cdots,8$) constructed from the two-point functions (\ref{eq:mock2pt}). The singular values are plotted in descending order.
One can see that $C^{+(m)}_{ij}$ has essentially zero singular values within double precision for $r\ge 6$; this makes sense because the number of states in our mock data is 6 and there are no further orthogonal states. In this case, the maximum possible low-rank approximation is at $r=5$.

%
% FIG.
%

We then truncate the Krylov subspace using $(r+1)$-highest singular values, and solve the TGEVP. In order to investigate systematic bias due to this low-rank approximation, we investigate the ground-state energy for various matrix sizes $m+1$ and for the low-rank approximation with $r$.

\begin{figure*}[tbp]
\centering
\includegraphics[width=0.7\textwidth,bb=0 30 745 550,clip]{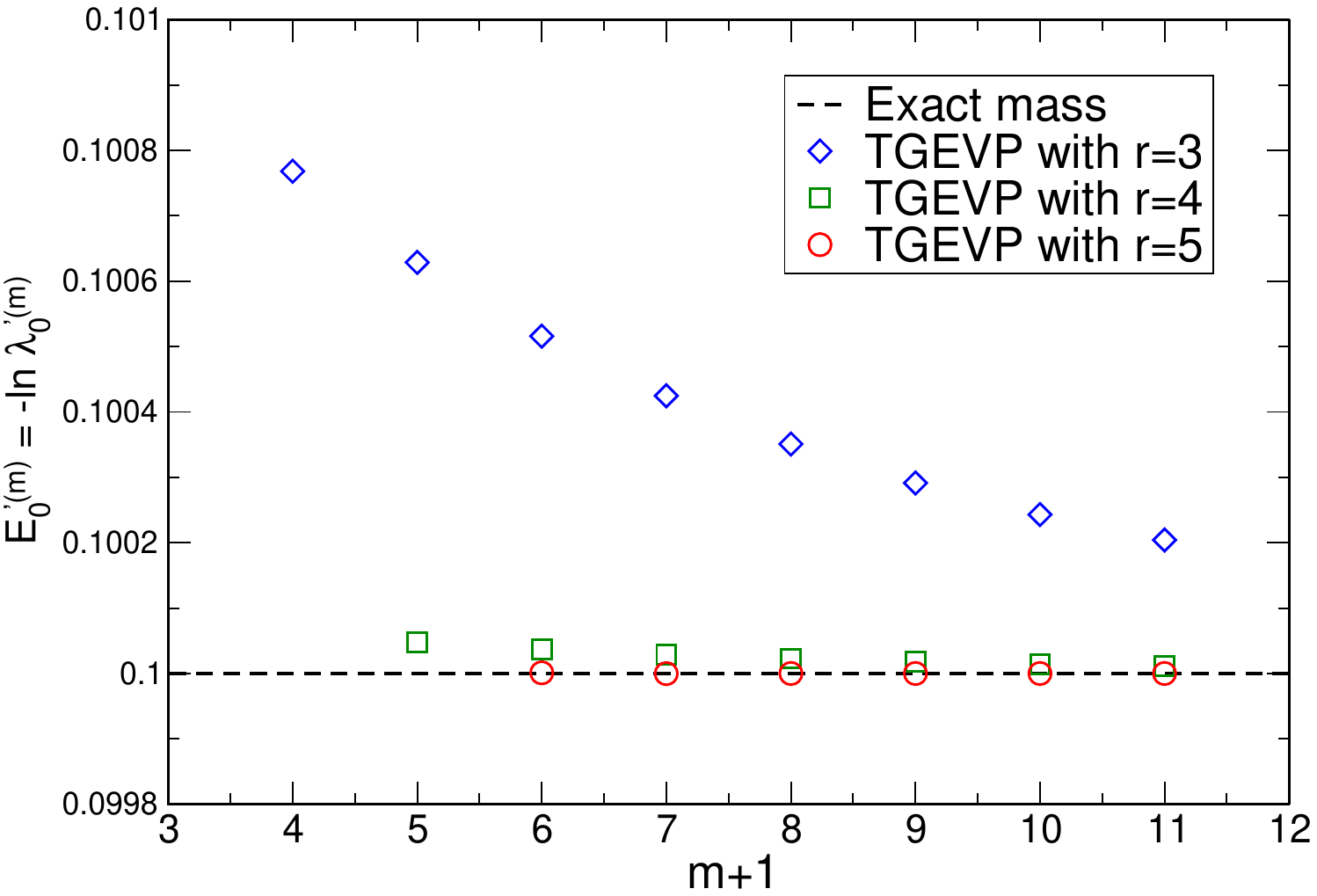}
\caption{
Estimated ground-state mass $E_0'^{(m)}=-\ln\lambda_0'^{(m)}$ for various $m$ and $r$.
The horizontal axis represents the matrix size $m+1$. The black dashed line shows the exact mass, which is an input when generating the mock data.
}
\label{fig:massNoiseless_TGEVP}
\end{figure*}

In Figure~\ref{fig:massNoiseless_TGEVP} we plot $E_0'^{(m)}=-\ln\lambda_0^{\prime(m)}$ as a function of the original matrix size $m+1$ for different choices of $r$. We confirmed that the largest eigenvalue evaluated by TGEVP with $r=5$ reproduces the input mass 0.10 to nine significant digits.
On the other hand, the eigenvalues evaluated with $r<5$ underestimate the eigenvalues as
$E_0'^{(8)}=-\mathrm{ln}\lambda_0^{\prime(8)}=0.100018$ ($r=4$) and
$E_0'^{(8)}=-\mathrm{ln}\lambda_0^{\prime(8)}=0.10029$ ($r=3$), showing bias due to the low-rank approximation. (We took $m=8$ as an example.)
We also observe that the bias decreases as the Krylov subspace $m+1$ is made large, even if the TGEVP is solved in a small truncated space of $r=3$.

%
% FIG.
%

\begin{figure*}[tbp]
\centering
\includegraphics[width=0.7\textwidth,bb=0 10 745 550,clip]{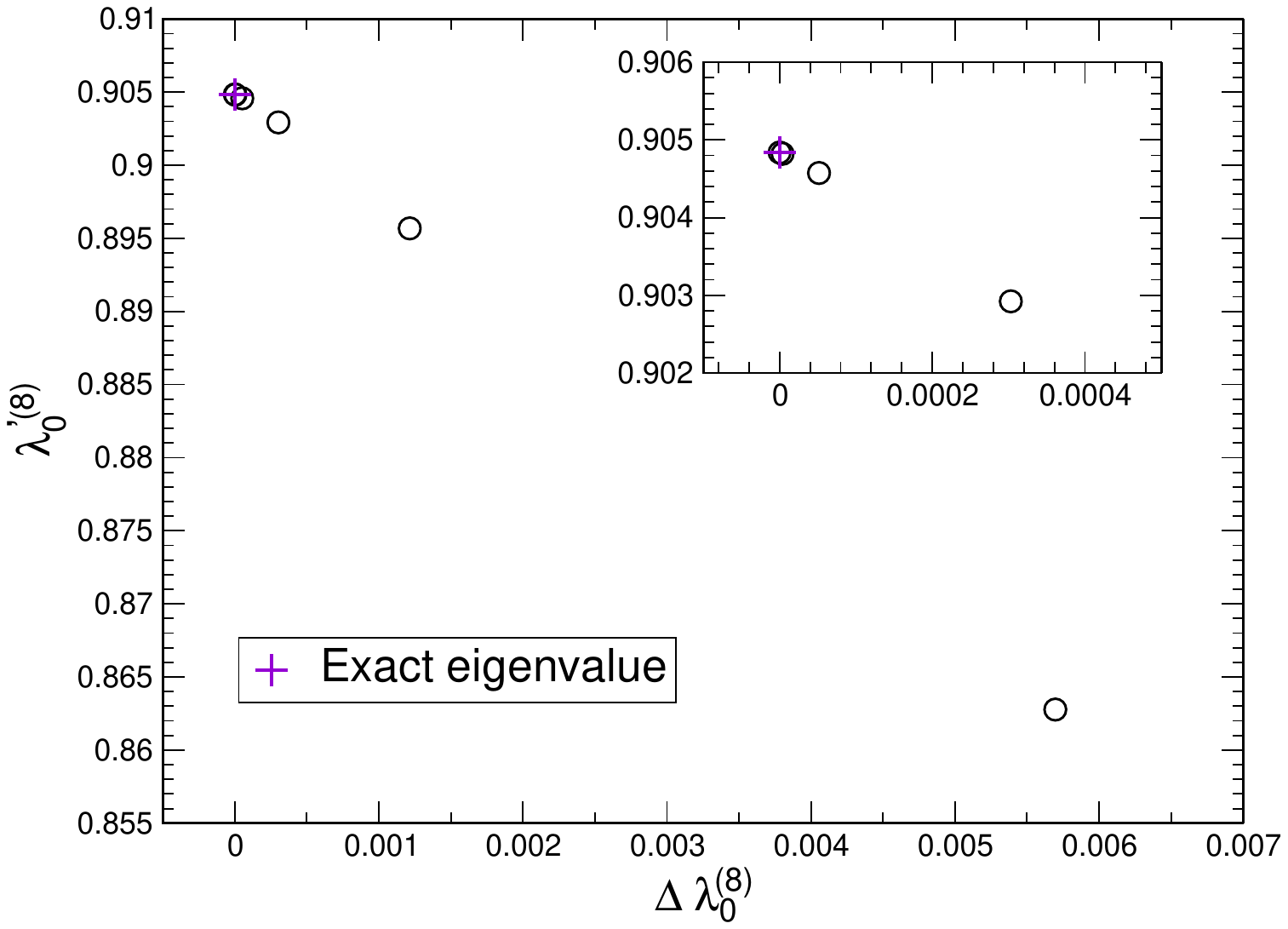}
\caption{
Largest eigenvalue $\lambda_0'^{(m)}$ plotted against the eigenvalue-variance $\Delta\lambda^{(m)}_0$. The base Krylov subspace is for $m=8$. Each point corresponds to $r=0,\cdots,5$ from right to left. The points for $r=4$ and 5 are nearly degenerate and only visible in the inset. The violet plus represents the exact mass.}
\label{fig:evNoiseless_extrapolation}
\end{figure*}

Now we introduce the eigenvalue-variance extrapolation. Figure~\ref{fig:evNoiseless_extrapolation} shows $\lambda^{\prime(8)}_0$ as a function of $\Delta\lambda^{(8)}_0$. We observe a nearly perfect linearity starting from the rightmost point of $r=0$, and a linear extrapolation with $r=3$ and $4$ to $\Delta\lambda^{(8)}_0=0$ gives $\tilde{\lambda}_0^{(8)}=0.90484$, which corresponds to $\tilde{E}_0'^{(8)}=-\mathrm{ln}{\tilde{\lambda}}_0^{(8)}=0.099996$, reproducing the input value almost exactly.

%
% FIG.
%

\begin{figure*}[tbp]
\centering
\includegraphics[width=0.7\textwidth,bb=0 10 745 550,clip]{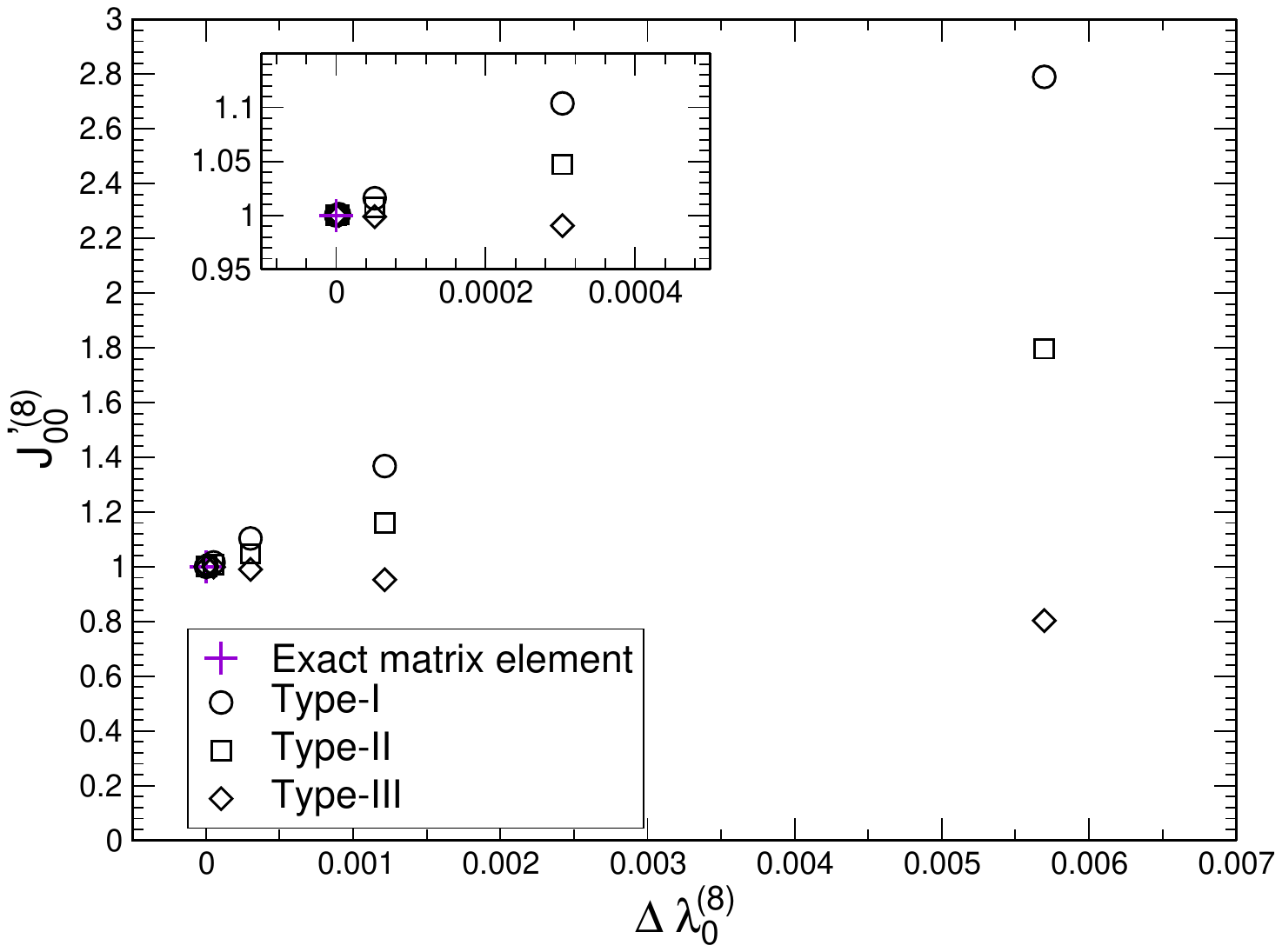}
\caption{
Ground-state matrix element $J_{00}'^{(8)}$ for three types of the three-point correlation function. The plot shows the dependence on the ``variance'' $\Delta \lambda^{(8)}_0$. From right to left, the points correspond to $=0,\cdots,5$.
Results with small $\Delta \lambda^{(m)}_0$ are also plotted in the inset. The violet plus represents the input value $J_{00}=1$.
}
\label{fig:meNoiseless_extrapolation}
\end{figure*}

The ground-state matrix element can also be estimated by using the low-rank eigenvectors as (\ref{eq:ground-state_matrix_element}). Figure~\ref{fig:meNoiseless_extrapolation} shows $J_{00}'^{(8)}$ as a function of $\Delta \lambda^{(8)}_0$ with three input types of the matrix elements: $J_{mn}^\mathrm{I}$, $J_{mn}^\mathrm{II}$, $J_{mn}^\mathrm{III}$. From right to left the points correspond to the different low-rank truncation $r=0,\cdots,5$. Again, the results approach 1.0, the input value, as $r$ is increased.

%
% FIG.
%

Recall that the leading effect of the bias due to the inexact eigenstate is proportional to the off-diagonal matrix element (\ref{eq:J00}), and it should depend on the ``variance'' as $\sqrt{\Delta\lambda_0^{(m)}}$. There is also a term linear in $\Delta\lambda_0^{(m)}$ so that the overall dependence could be complicated. However, the plot in Figure~\ref{fig:meNoiseless_extrapolation} shows an almost linear dependence. If we extrapolate linearly from the furthest two points ($r$=0 and 1), the results are $\tilde{J}_{00}^{\mathrm{I}(8)}=0.9826$, $\tilde{J}_{00}^{\mathrm{II}(8)}=0.9880$, and $\tilde{J}_{00}^{\mathrm{III}(8)}=0.9934$. They reproduce the input to 1--2\%, suggesting that the effects of the square-root term is not very significant. Still, the deviation from the exact result depends on the size of the off-diagonal components, {\it i.e.} three models among which ``I'' has the most significant off-diagonal terms while ``III'' does not have the off-diagonal matrix element. If we extrapolate with $r=3$ and 4, the deviation from the exact result is at the level of $10^{-5}$ for all three models.

\subsection{Noisy mock data}
\label{ssec:noisy_mock_data}

Next, we consider the estimation of the ground-state mass and matrix element with noisy data sets. 
We add a randomly generated noise using the normal distribution $\mathcal{N}(\mu,\sigma)$, where $\mu$ and $\sigma$ represent mean and variance. The noisy mock data of $C(t)$ is created as $C_\mathrm{noisy}(t) = C(t) + \epsilon(t)$, where the noise term is sampled from a distribution as $\epsilon(t)=C(t)\times \mathcal{N}(0,1)/100$, so that the fluctuation is at the level of 1\% of the data. We create 500 statistical samples. The three-point function $C^{3\mathrm{pt}}(t_\mathrm{sep},t)$ is provided as 
$C^{3\mathrm{pt}}_{\mathrm{noisy}}(t_\mathrm{sep},t) = C^{3\mathrm{pt}}(t_\mathrm{sep},t) + \epsilon^{3\mathrm{pt}}(t_\mathrm{sep})$
with $\epsilon^{3\mathrm{pt}}(t_\mathrm{sep})=C^{3\mathrm{pt}}(t_\mathrm{sep},t_\mathrm{sep})\times \mathcal{N}(0,1)/100$.
We create $500$ bootstrap samples, in which each bootstrap sample is constructed by randomly sampling from $500$ configurations with replacement. 

\begin{figure*}[tbp]
\centering
\includegraphics[width=0.7\textwidth,bb=0 30 745 550,clip]{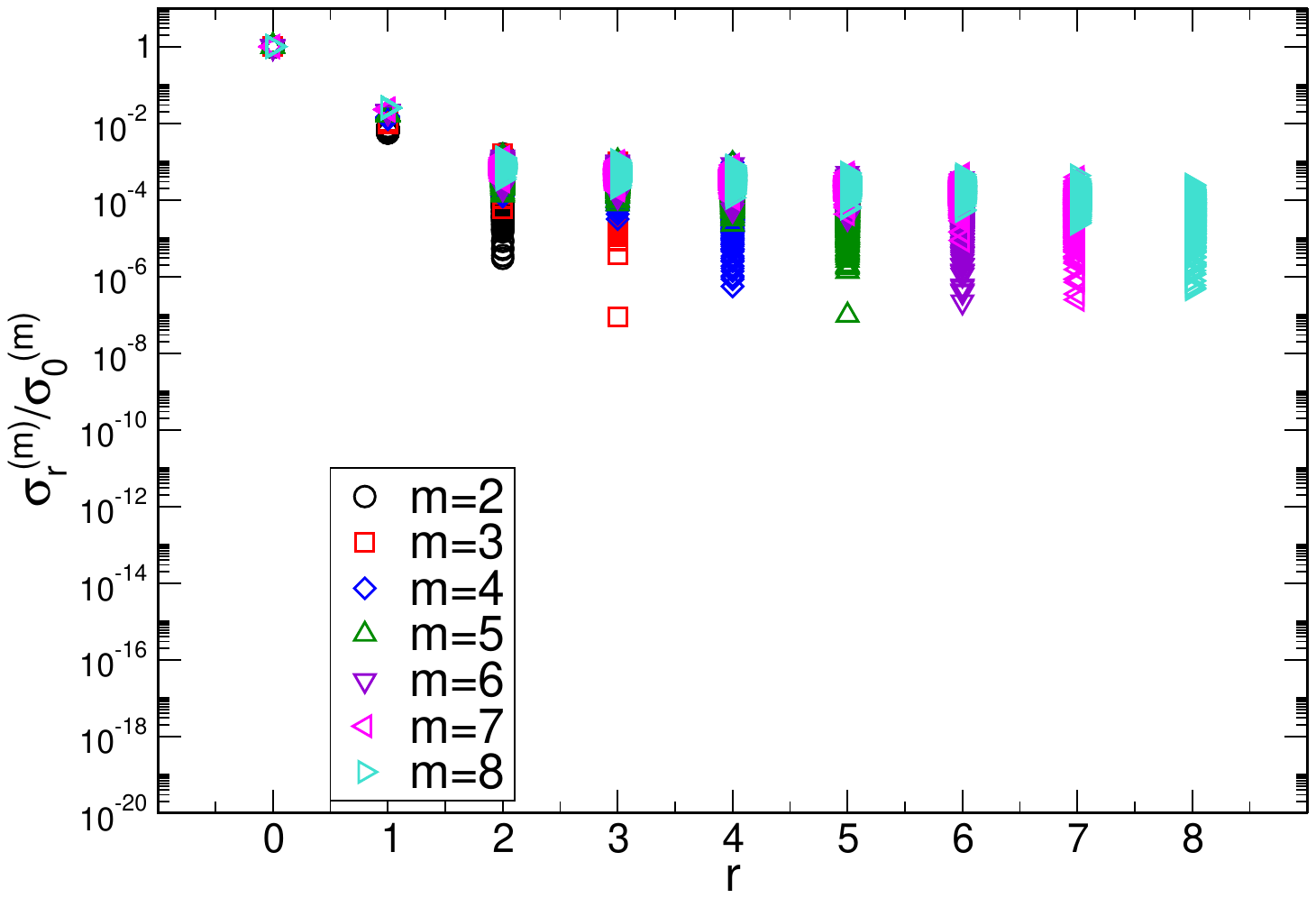}
\caption{
Same as Fig.~\ref{fig:svNoiseless} but with noisy data sets. Each symbol represents a result for a given bootstrap sample; there are 500 of them nearly overlapping in the plot.
}
\label{fig:svNoisy}
\end{figure*}

We apply the SVD and low-rank approximation for each bootstrap sample before solving the TGEVP. Figure~\ref{fig:svNoisy} represents the normalized singular values $\sigma_r^{(m)}/\sigma_0^{(m)}$ of the $(m+1)\times(m+1)$ matrix $C^{+(m)}_{ij}$ defined in Eq.~(\ref{eq:matrix_C+}) for all bootstrap samples of noisy two-point correlation functions. In the plot, the result for each bootstrap sample is shown; they are distributed in some range. With the size of the statistical fluctuation introduced for these mock data, the distribution of the resulting singular values overlaps for $r\ge2$. This means that the associated eigenstates would not be definitely identified for a given $r$ greater than 2, and we can use only the $r\le 1$ subspace to define the low-rank approximation.
%
% FIG.
%

%
% FIG.
%
\begin{figure*}[tbp]
\centering
\includegraphics[width=0.7\textwidth,bb=0 30 745 550,clip]{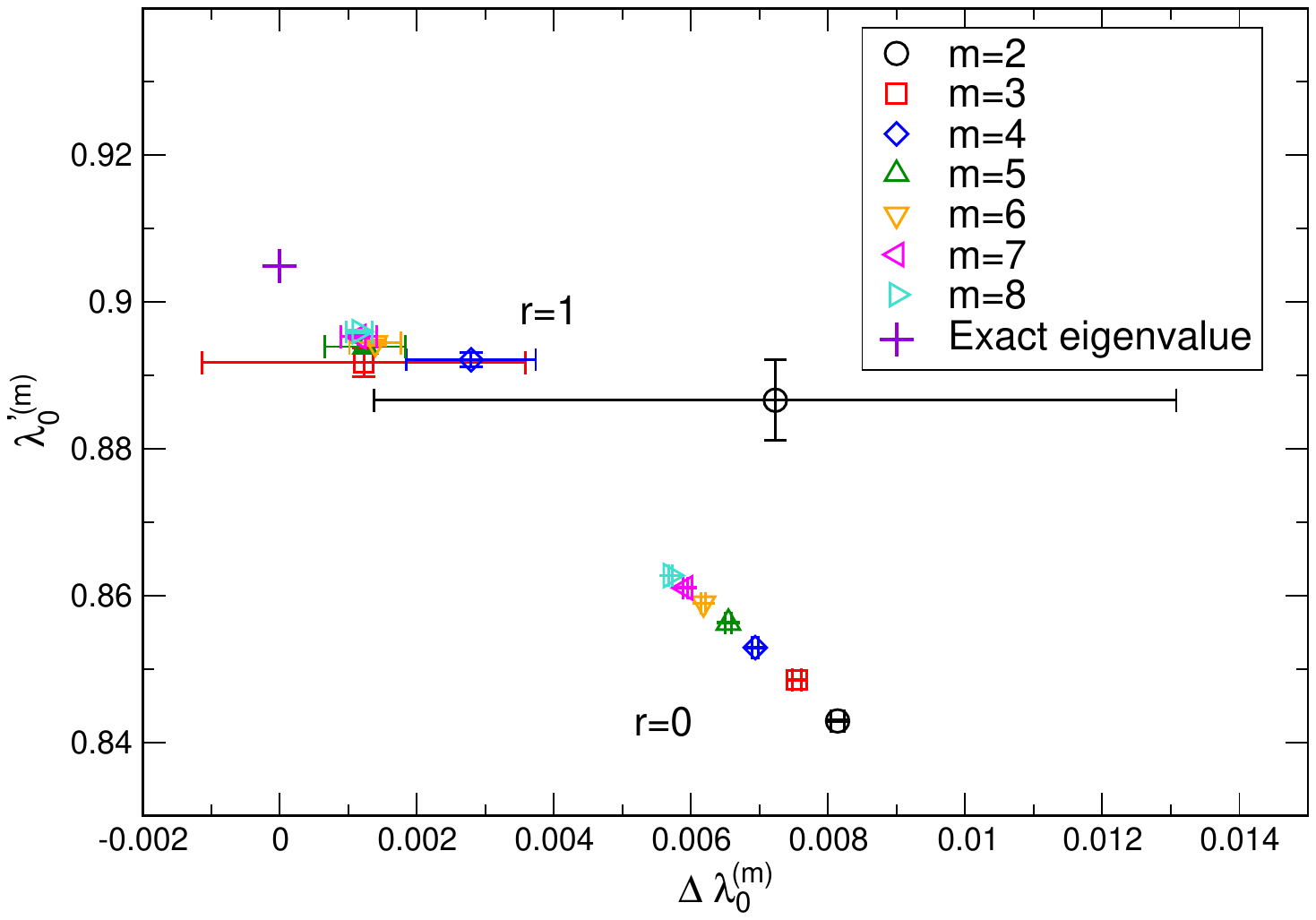}
\caption{
    Largest eigenvalue $\lambda_0'$ as a function of $\Delta\lambda^{(m)}_0$.
    The results with $r=0$ and 1 are shown for $m=2,\cdots,8$. The error bar is estimated by the bootstrap analysis. The violet plus represents the input.
}
\label{fig:evNoisy_extrapolation}
\end{figure*}

Figure~\ref{fig:evNoisy_extrapolation} shows the highest eigenvalue $\lambda^{\prime(m)}_0$ obtained with $r=0$ and 1 as a function of $\Delta\lambda_0^{(m)}$ for various values of $m$ $=2,\cdots,8$. We observe that the results are closer to the input value (shown by a plus symbol) with $r=1$ albeit larger statistical error. Thus, a linear extrapolation towards $\Delta\lambda_0^{(m)}\to 0$ with these data would lead to a more accurate estimate of the eigenvalue even with noise. The $m$-dependence of the slope is not significant.

%
% FIG.
%
\begin{figure*}[tbp]
\centering
\includegraphics[width=0.7\textwidth,bb=0 30 745 550,clip]{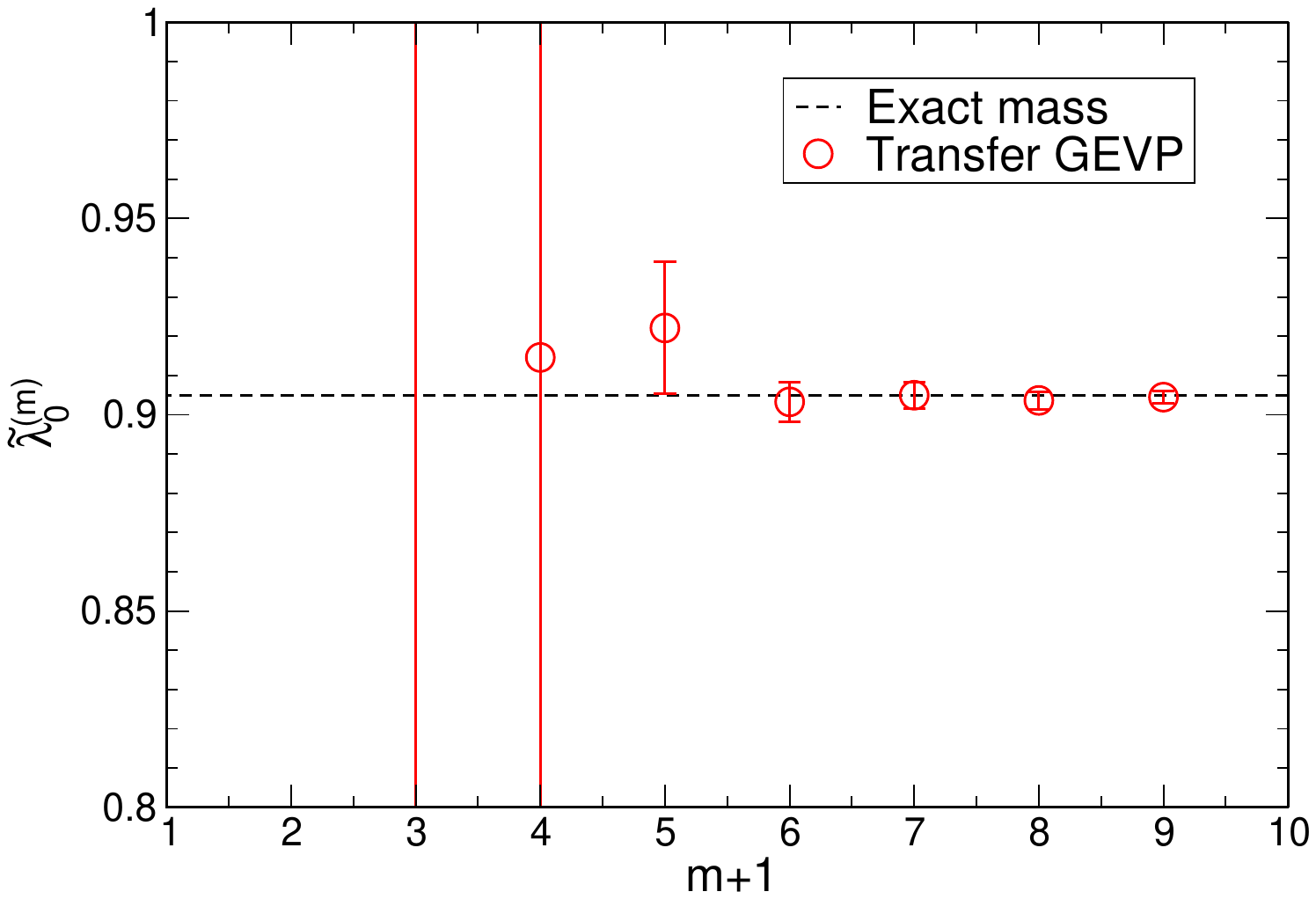}
\caption{
    Estimate of the largeset eigenvalue obtained by an extrapolation $\tilde{\lambda}_0^{(m)}$ with the data at $r=0$ and 1. The results are shown for various $m+1$. The input value is shown by a horizontal dashed line.
}
\label{fig:evNoisy_bootstrap}
\end{figure*}

Figure~\ref{fig:evNoisy_bootstrap} shows the results of the extrapolation $\tilde{\lambda}_0^{(m)}$. With sufficiently large $m+1$ ($\ge 5$) we reproduce the input value. Numerically, they are $\tilde{\lambda}^{(8)}_0=0.9045(16)$ and $\tilde{E}_0^{(8)}=-\mathrm{ln}{\tilde{\lambda}}^{(8)}_0 = 0.1004(18)$. This indicates that our strategy successfully estimates the exact eigenvalue within the statistical uncertainty, as far as the original Krylov subspace $m+1$ is sufficiently large.

\begin{figure*}[tb]
\centering
\includegraphics[width=0.8\textwidth,bb=10 10 725 430,clip]{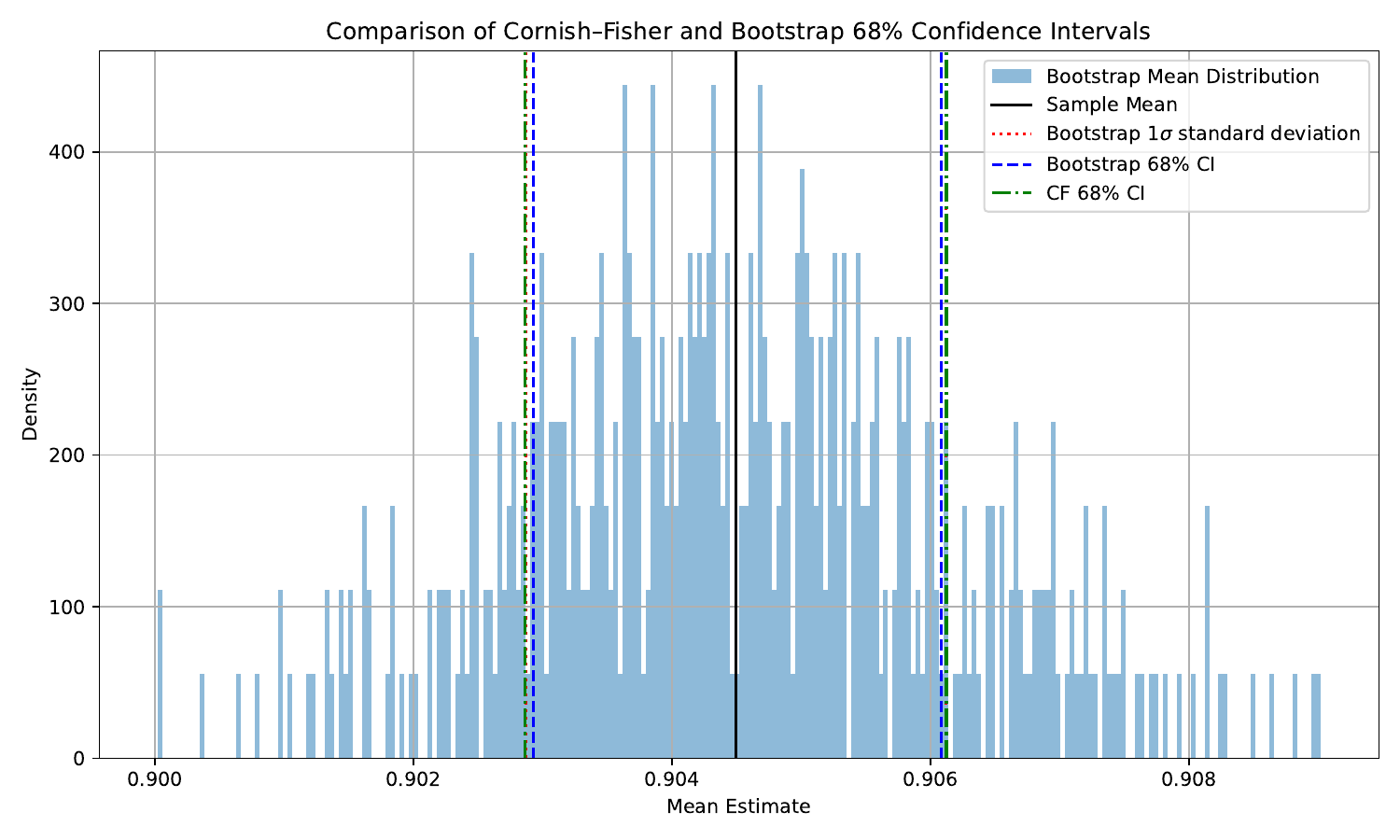}\\
\caption{
    Histogram of $\tilde{\lambda}_0^{(m=8)}$.
    The statistical uncertainty is evaluated by three approaches:
    the 1$\sigma$ standard deviation (red dotted), the 68\% percentile interval (blue dashed), and the Cornish-Fisher corrected 68\% CI (green dot-dashed).
    The black solid line represents the mean of the extrapolated value.
}
\label{fig:evmassNoisy_distribution}
\end{figure*}

The statistical error of the estimated eigenvalue represents that of one standard deviation assuming the normal distribution.
Figure~\ref{fig:evmassNoisy_distribution} shows the distribution of $\tilde{\lambda}_0^{(8)}$'s from each bootstrap sample. We observe that the distribution resembles the normal distribution. Indeed, a standard deviation assuming the normal distribution (red dotted) agrees well with the 68\% percentile interval (blue dashed) and the Cornish-Fisher corrected 68\% confidence interval (green dot-dashed).
(See Appendix~\ref{app:bootstrap_resampling_to_evaluate_statistical_uncertainty} for the Cornish-Fisher prescription.)
If the results were not normally distributed, those two intervals would be different from a standard deviation. We therefore conclude that spurious eigenvalues are filtered out by our strategy and that the central limit theorem with bootstrap resampling does not break down.

%
% FIG.
%

\begin{figure*}[tbp]
\centering
\includegraphics[width=0.7\textwidth,bb=10 10 475 350,clip]{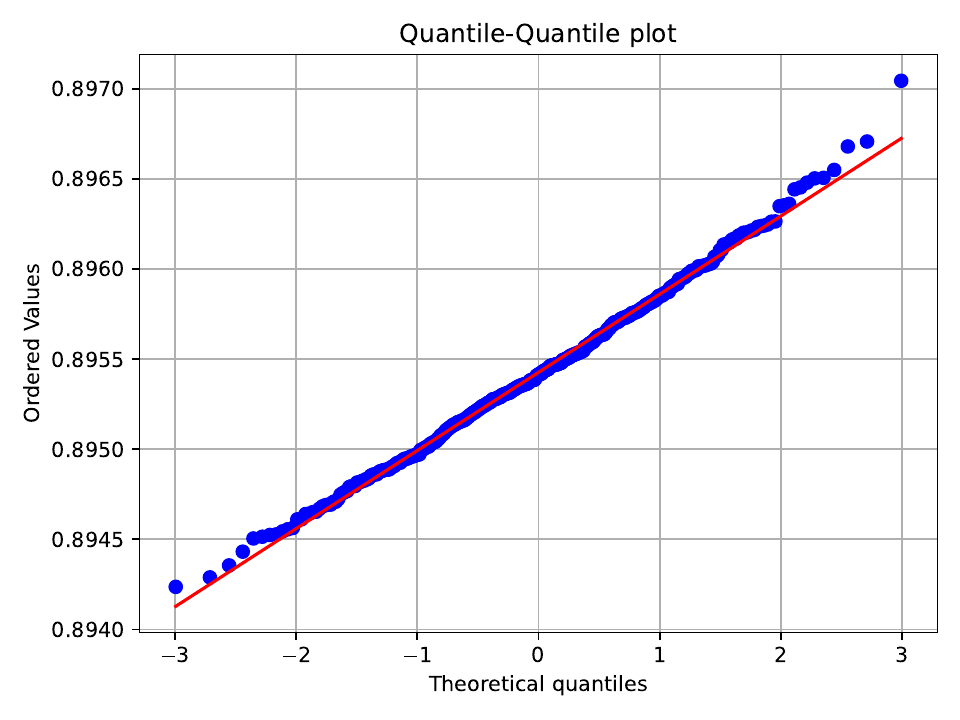}
\includegraphics[width=0.7\textwidth,bb=10 10 475 350,clip]{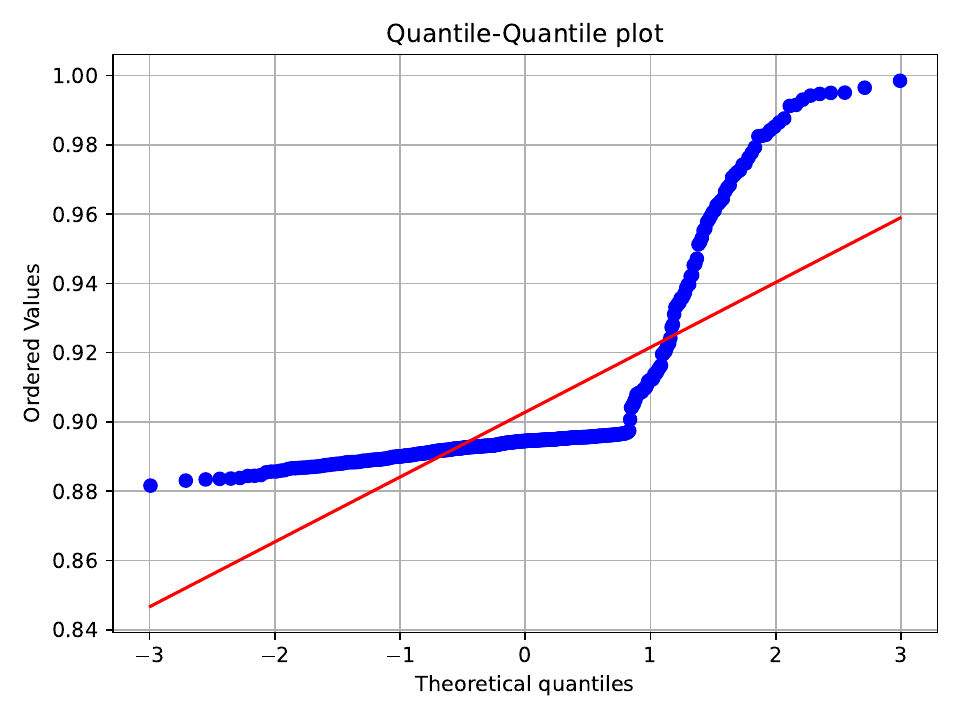}
\caption{
Quantile-Quantile plot for the distribution of the ground-state eigenvalues obtained with $m=8$ for $r=1$ (top) and $r=2$ (bottom). The horizontal axis represents the quantiles of the sampled eigenvalues, and the vertical axis shows the corresponding eigenvalue. The blue dots are obtained from the sampled eigenvalues, while the red solid line is the result with a least-square regression with the normal distribution.
}
\label{fig:qqplot}
\end{figure*}

It is instructive to see how the statistical distribution is distorted when the spurious eigenvalues are included in the analysis. For this purpose, we introduce the Quantile-Quantile (QQ) plot, which displays the quantiles of the dataset compared to the quantiles of the expected theoretical distribution (the normal distribution in our case).
Figure~\ref{fig:qqplot} represents the QQ plot for the distribution of the ground-state eigenvalue obtained with $m=8$ for $r=1$ (top) and 2 (bottom). If the data followed the normal distribution, these points would have been arranged in a straight line. For data with $r=1$, we observe that the data points fall in a straight line, which means that the distribution of the eigenvalues is almost identical to the normal distribution. On the other hand, data points with $r=2$ show a deviation in the tail, indicating that the data have a heavier tail than the normal distribution.

%
% FIG.
%

\begin{figure*}[tb]
\centering
\includegraphics[width=0.7\textwidth,bb=10 10 725 430,clip]{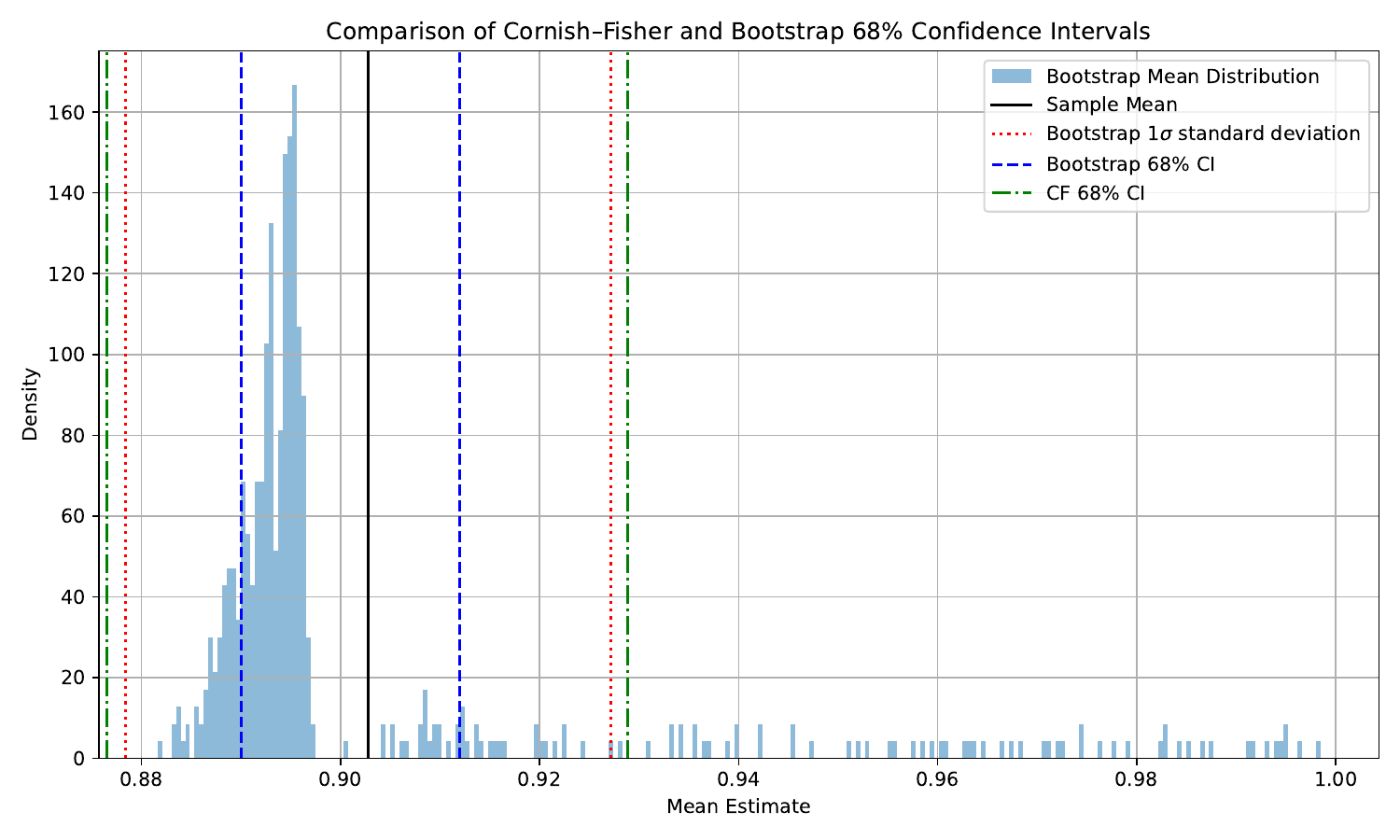}
\caption{
Distribution of the ground-state eigenvalues and a comparison of three approaches to estimate the statistical uncertainty: the 1$\sigma$ standard deviation ($\delta \hat{\theta}^*$, red dotted), the 68\% percentile interval ($\mathrm{CI_{bootstrap}}$, blue dashed), and the Cornish-Fisher corrected 68\% CI ($\mathrm{CI_{CF}}$, green dot-dashed).
The black solid line represents the sample mean.
The data are obtained with $m=8$ for $r=2$.
}
\label{fig:evNoisy_distribution}
\end{figure*}

The suspicious case $r=2$ can also be seen using the histogram.
Figure~\ref{fig:evNoisy_distribution} shows the eigenvalue distribution for $r=2$, clearly demonstrating a deviation from the normal distribution. The estimates of the error from the standard deviation and those of the 68\% confidence interval, as well as its Cornish-Fisher variant, are largely different. 

%
% FIG.
%

\begin{figure*}[tbp]
\centering
\includegraphics[width=0.52\textwidth,bb=20 30 745 550,clip]{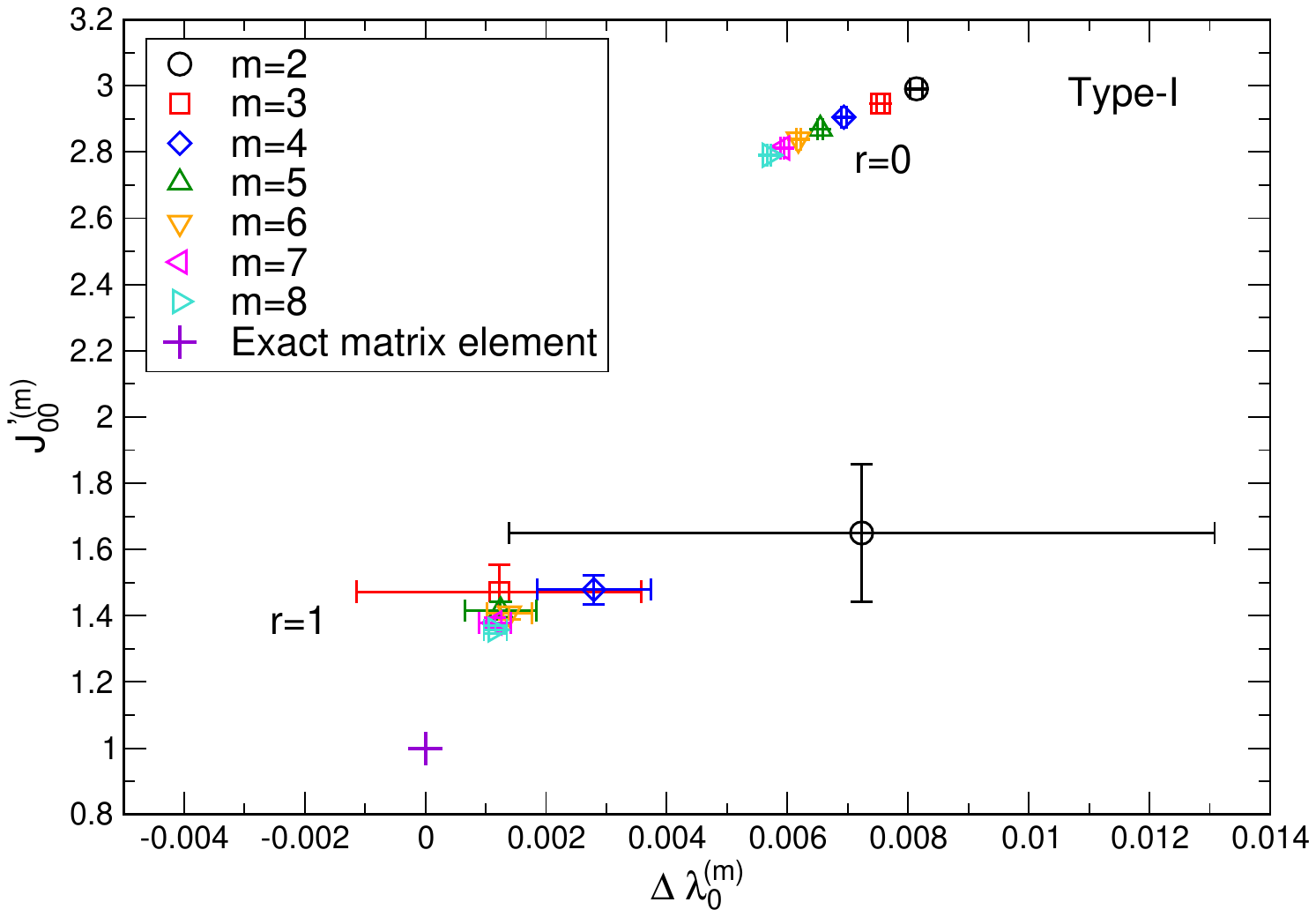}\\
\includegraphics[width=0.52\textwidth,bb=20 30 745 550,clip]{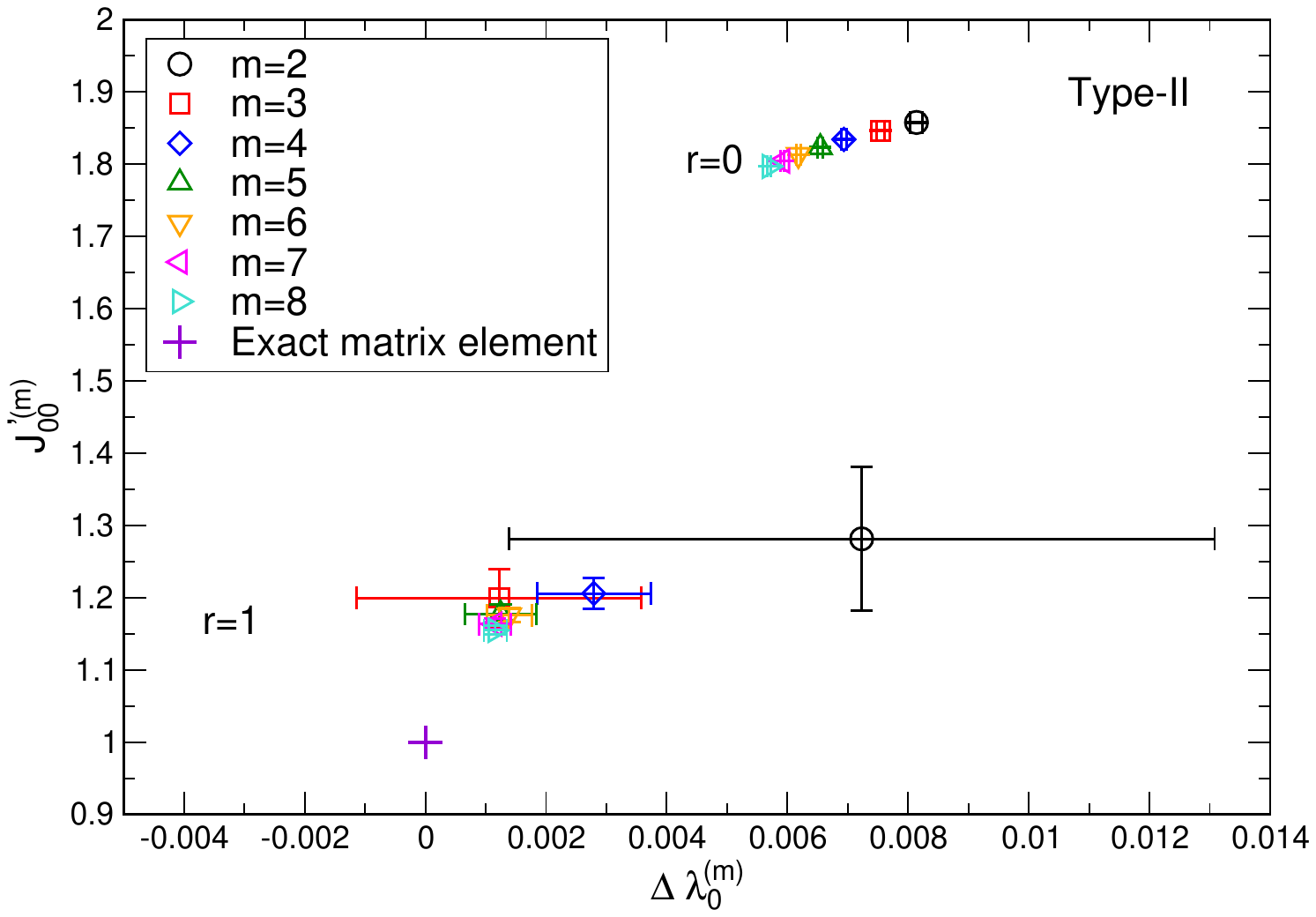}\\
\includegraphics[width=0.52\textwidth,bb=20 30 745 550,clip]{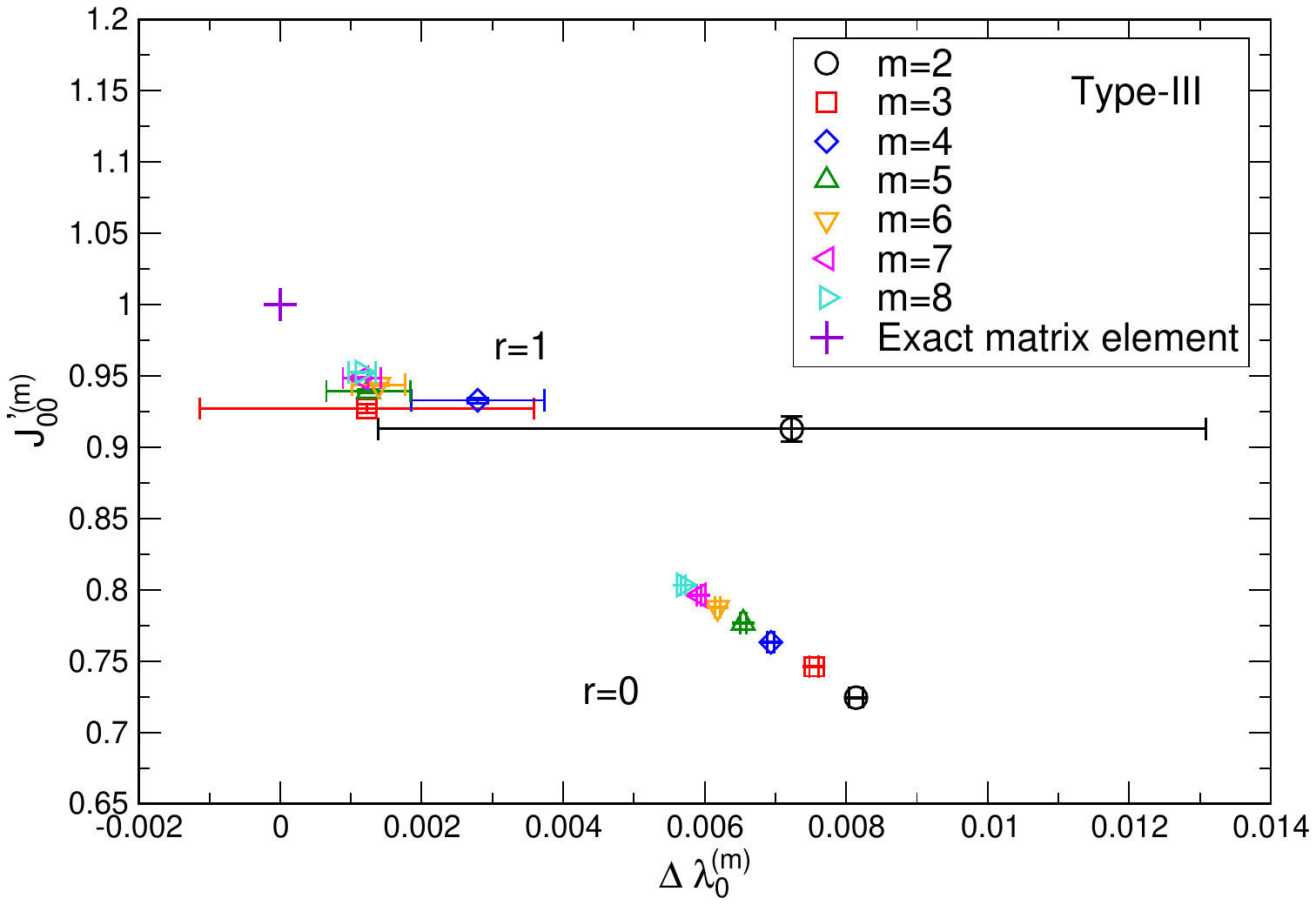}
\caption{
    Ground-state matrix element $J_{00}'^{(m)}$ from the mock data with noise. The input is type-I (top), II (middle) and III (bottom). The results with the low-rank approximation with $r=0$ and 1 are shown as a function of $\Delta\lambda^{(m)}_0$. The violet plus represents the input.
}
\label{fig:mePTNoisy_extrapolation}
\end{figure*}

\begin{figure*}[tbp]
\centering
\includegraphics[width=0.52\textwidth,bb=0 30 745 550,clip]{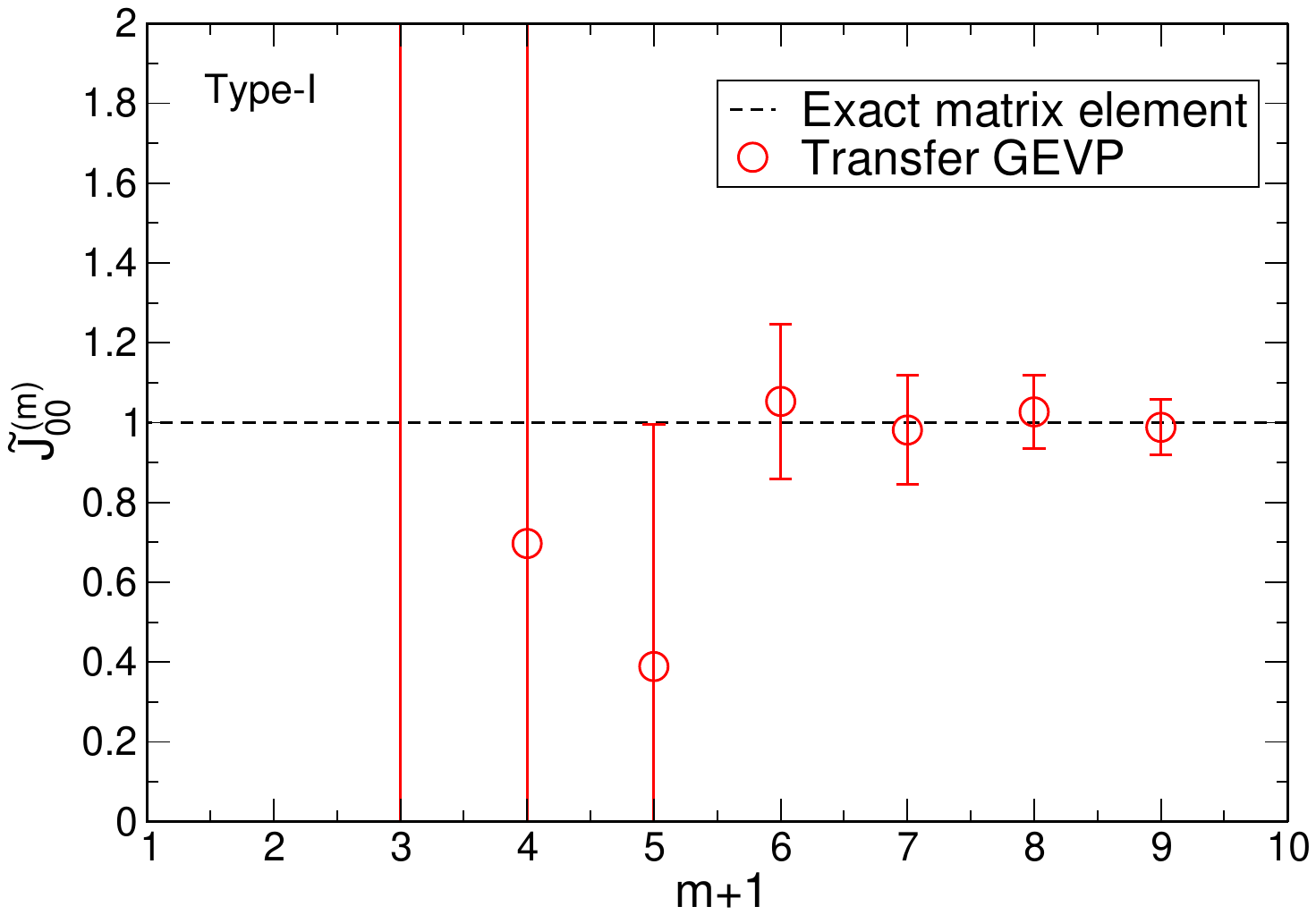}\\
\includegraphics[width=0.52\textwidth,bb=0 30 745 550,clip]{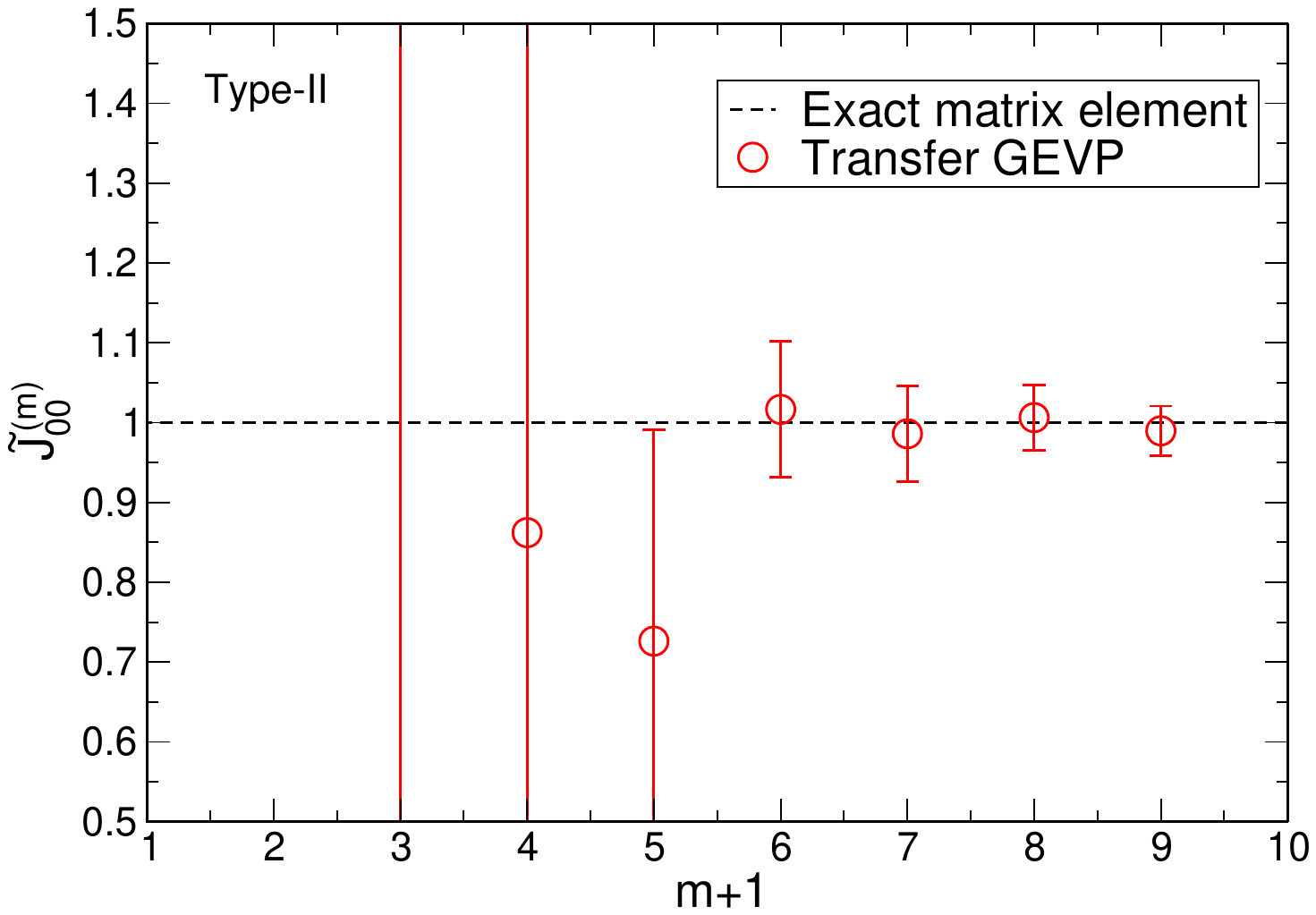}\\
\includegraphics[width=0.52\textwidth,bb=0 30 745 550,clip]{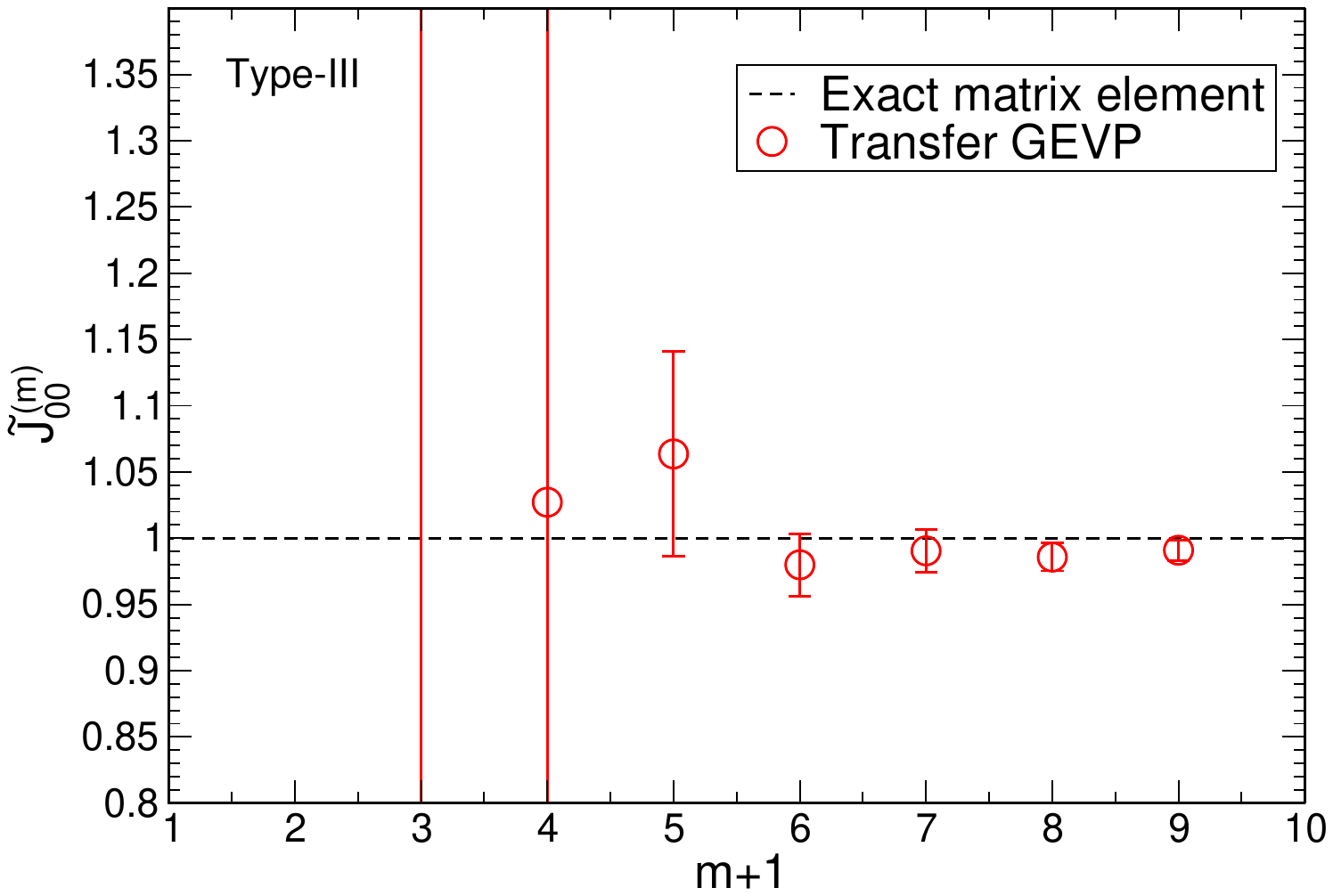}
\caption{
    Ground-state matrix element $J_{00}'^{(m)}$ from the mock data with noise. The results are extrapolated to $\Delta\lambda^{(m)}_0\to 0$. The input is type-I (top), II (middle) and III (bottom).
}
\label{fig:mePTNoisy_bootstrap}
\end{figure*}

Regarding the estimation of the ground-state matrix element, Figure~\ref{fig:mePTNoisy_extrapolation} shows the matrix elements for three input choices. They are plotted as functions of $\Delta\lambda_0^{(m)}$. 
The size of the Krylov subspace is $m=2,\cdots,8$ and $r=0,1$. Like the eigenvalue, the matrix elements show the trend to approach the input value as we expand the rank from $r=0$ to 1; the extrapolation to $\Delta\lambda_0^{(m)}\to 0$ further improves the accuracy. The results of the extrapolation with $r=0$ and 1 are shown in Figure~\ref{fig:mePTNoisy_bootstrap} for various $m$. Beyond $m+1\ge 6$ we reproduce the input within the statistical error. The estimate of the statistical error with one standard deviation agrees well with those from the 68\% percentile interval or the Cornish-Fisher corrected one.

%
% FIG.
%
%
%
% FIG.
%

%%%%%%%%%%%%%%  SEC 5  %%%%%%%%%%%%%%%%%%%%%%%%
\section{Application to realistic lattice data}
\label{sec:application_to_realistic_lattice_data}
We apply the strategy to extract ground-state mass and matrix element to $K$ and $D_s$ meson correlators generated on 2+1-flavor M\"obius domain-wall fermion ensembles. The lattice spacing is $a\simeq0.080$ fm [$a^{-1}$ = 2.453(44)~GeV] and the lattice size is $32^3\times 64$. The pion mass is about 230~MeV. Other details of the ensemble are described in the supplementary material of \cite{Colquhoun:2022atw}.

We use the M\"obius domain-wall fermion for light, strange and charm quarks, and set the strange and charm quark masses close to their physical values. We computed the two-point correlators $C_P^{LL}(t)$ on 100 gauge configurations with a single choice of the source point on each configuration. 
The two-point correlators are generated with pseudo-scalar local operators of quantum numbers corresponding to $K$ and $D_s$ mesons. The spatial momentum is set to zero. We also consider a three-point correlation function corresponding to the $D_s\to D_s$ process with an insertion of temporal vector current $V_4^{lat}=\bar{c}\gamma_4 c$. Only the connected quark-line contraction is considered. To apply the TGEVP method, we prepare the data of $2\le t_\mathrm{sep}\le14$.

The local vector current on the lattice receives a finite renormalization.
The renormalization factor is included in the definition of the current as $V_4=Z_V V_4^{lat}$, where the value of $Z_V$ is taken from \cite{Tomii:2016xiv}: $Z_V=0.9553(53)$.

In order to evaluate the mean values and statistical uncertainties, we perform bootstrap resampling to create 500 bootstrap samples.

%%%%%%%%%%%%%%  SEC 6  %%%%%%%%%%%%%%%%%%%%%%%%

\subsection{Mass of the ground state}
\label{ssec:mass_of_the_groun_state}

Following Section~\ref{sec:numerical_test_with_mock_data}, we take $t_0=1$ in the lattice unit and analyze the normalized two-point correlation functions $\bar{C}_P^{LL}(t)=C_P^{LL}(t+2t_0)/C_P^{LL}(2t_0)$.

\begin{figure*}[tb]
\centering
\includegraphics[width=0.7\textwidth,bb=0 30 745 550,clip]{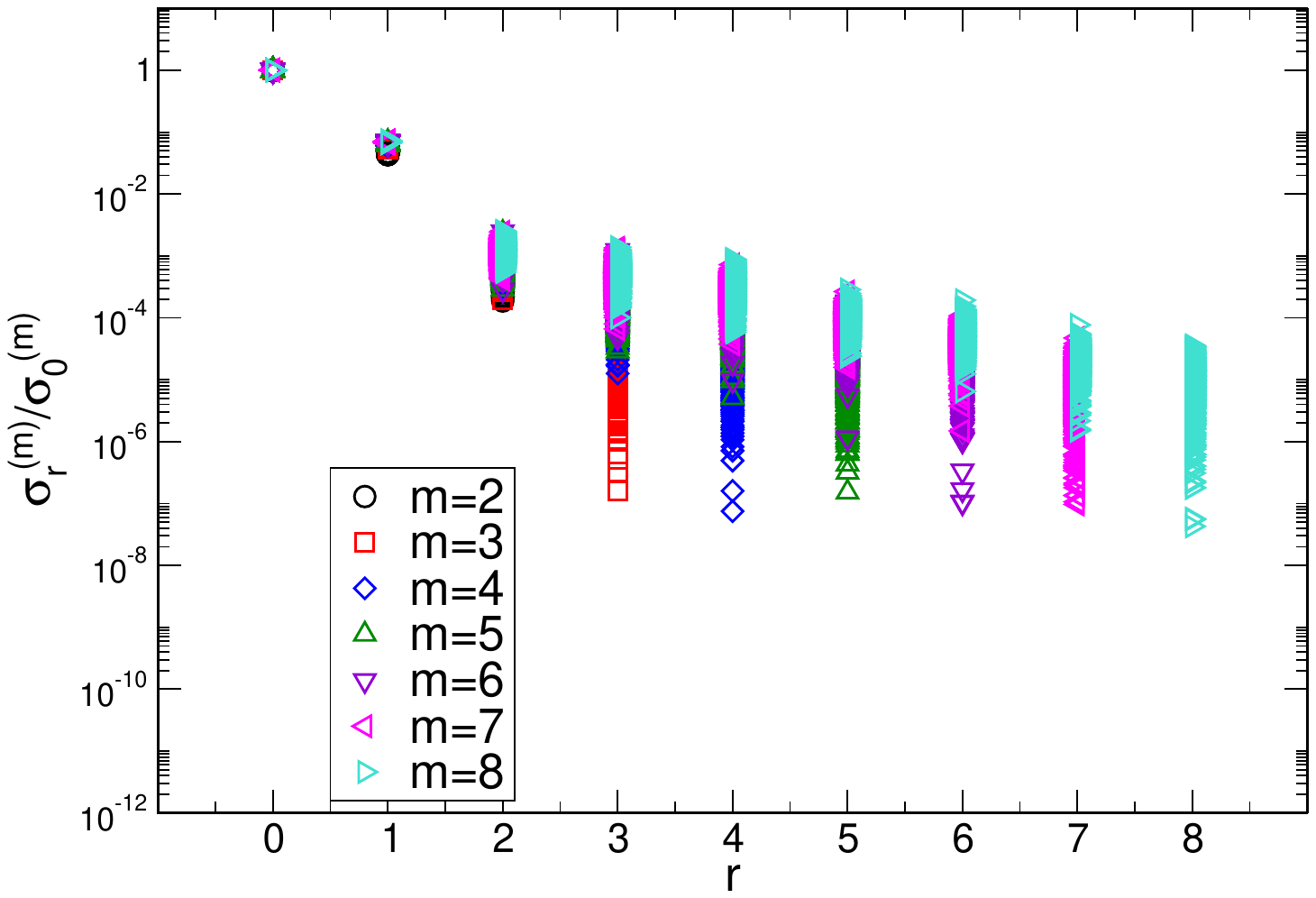}
\includegraphics[width=0.7\textwidth,bb=0 30 745 550,clip]{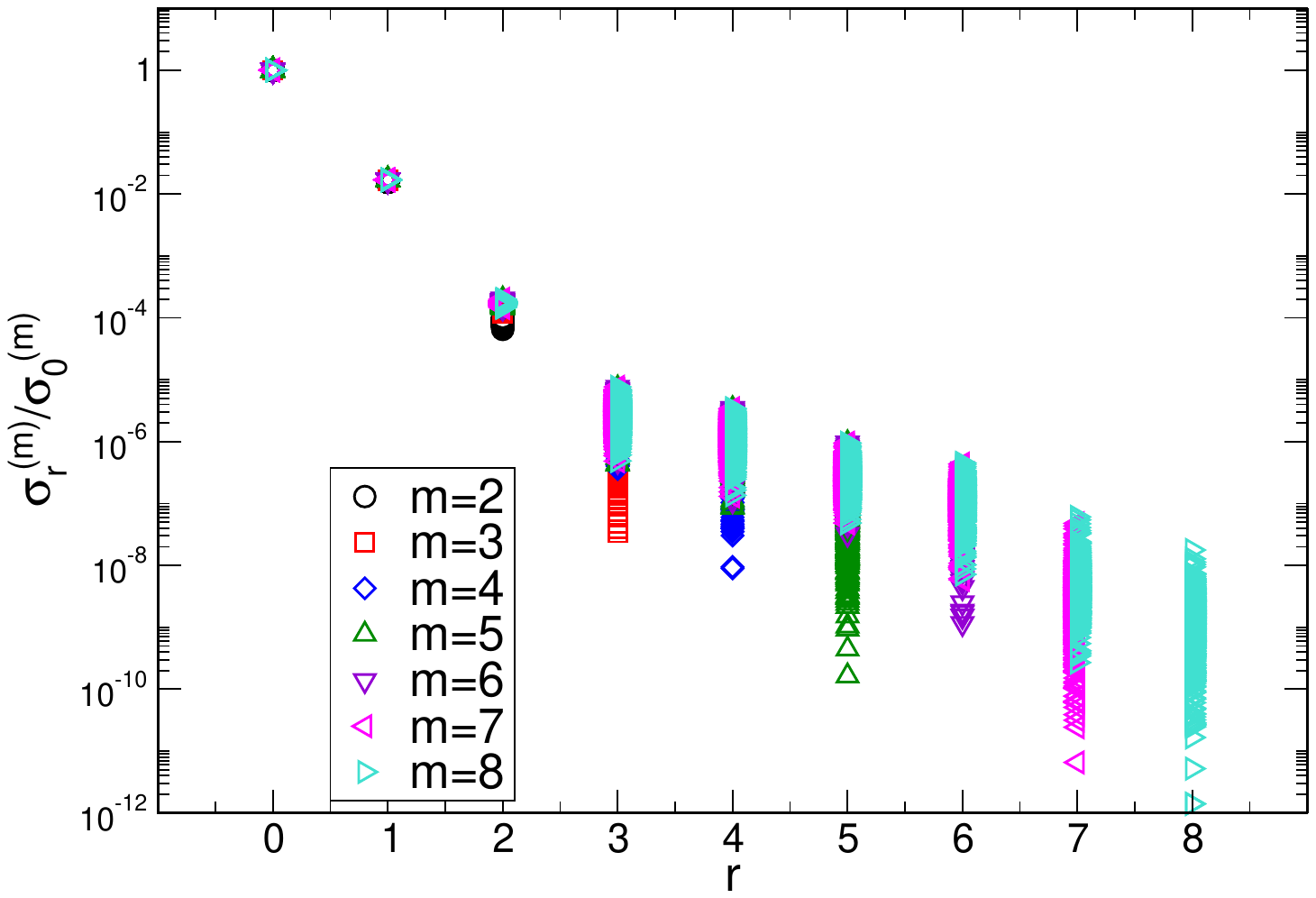}
\caption{
Normalized singular values of the matrix $C^{+(m)}_{ij}$ constructed from the $K$ (top panel) and $D_s$ (bottom) meson two-point correlation functions. Results for each bootstrap sample are plotted. 
The horizontal axis $r$ labels the singular values ordered in a descending order.
}
\label{fig:svLL_bootstrap}
\end{figure*}

Figure~\ref{fig:svLL_bootstrap} shows the normalized sigular values $\sigma_r^{(m)}/\sigma_r^{(0)}$ of the $(m+1)\times(m+1)$ matrix $C^{+(m)}_{ij}$, Eq.~(\ref{eq:matrix_C+}), created from $C_P^{LL}(t)$. The results for each bootstrap sample are plotted for the $K$ and $D_s$ channels. We observe that the distribution of the singular values of $r\ge 2$ overlaps for the $K$ (top panel) and of $r\ge 3$ for the $D_s$ (bottom). This implies that we can only use $r\le 1$ and $r\le 2$ subspaces to perform the low-rank approximation for the $K$ and $D_s$ channels, respectively.

%
% FIG.
%

\begin{figure*}[tb]
\centering
\includegraphics[width=0.7\textwidth,bb=20 30 745 550,clip]{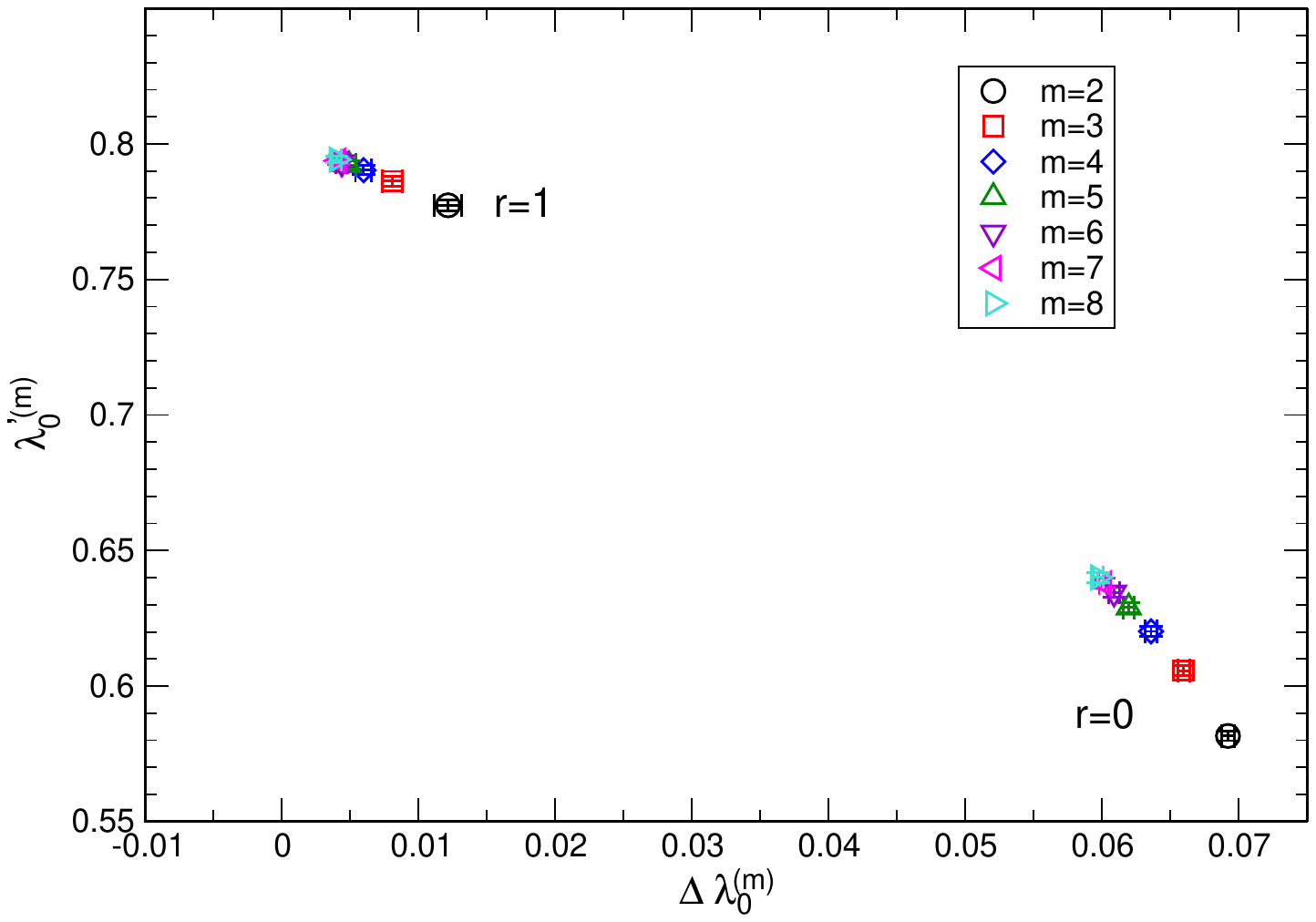}
\includegraphics[width=0.7\textwidth,bb=20 30 745 550,clip]{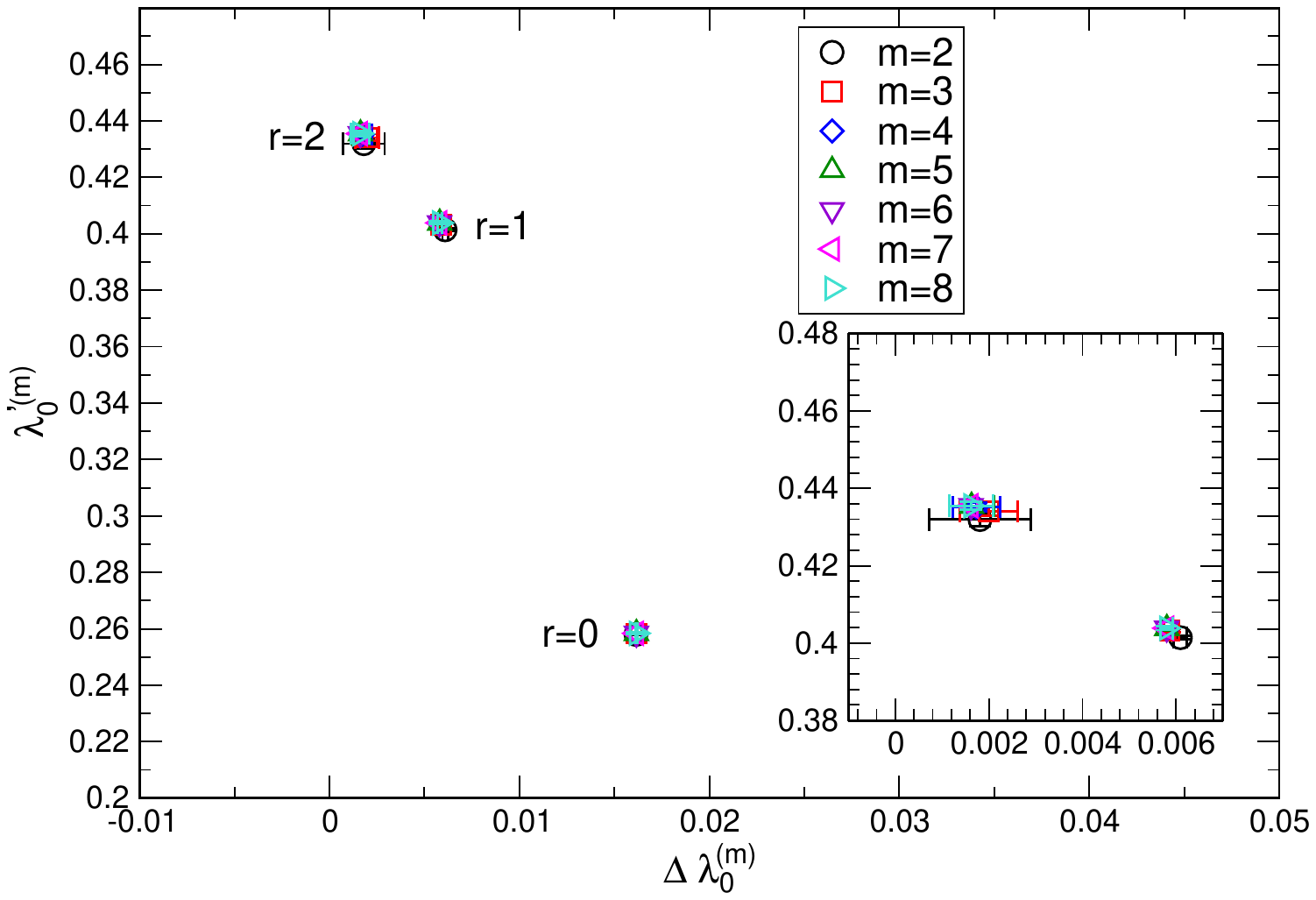}
\caption{
    Eigenvalue $\lambda_0'^{(m)}$ as a function of $\Delta\lambda^{(m)}_0$. The data for $K$ (top panel) and $D_s$ (bottom). The results from various $m$ with $r\le 1$ ($K$) or $r\le 2$ ($D_s$) are plotted. The inset magnifies the points of $r=2$ and 1.
}
\label{fig:evLL_s-l_DLbootstrap}
\end{figure*}

In Figure~\ref{fig:evLL_s-l_DLbootstrap} we plot the eigenvalues $\lambda_0^{\prime(m)}$ obtained in the low-rank approximation with $r\le 1$ for $K$ (top panel) and $r\le 2$ for $D_s$ (bottom). They are plotted as functions of  $\Delta\lambda^{(m)}_0$ for $m=2,\cdots,8$.

\begin{figure*}[tbp]
\centering
\includegraphics[width=0.7\textwidth,bb=0 10 745 550,clip]{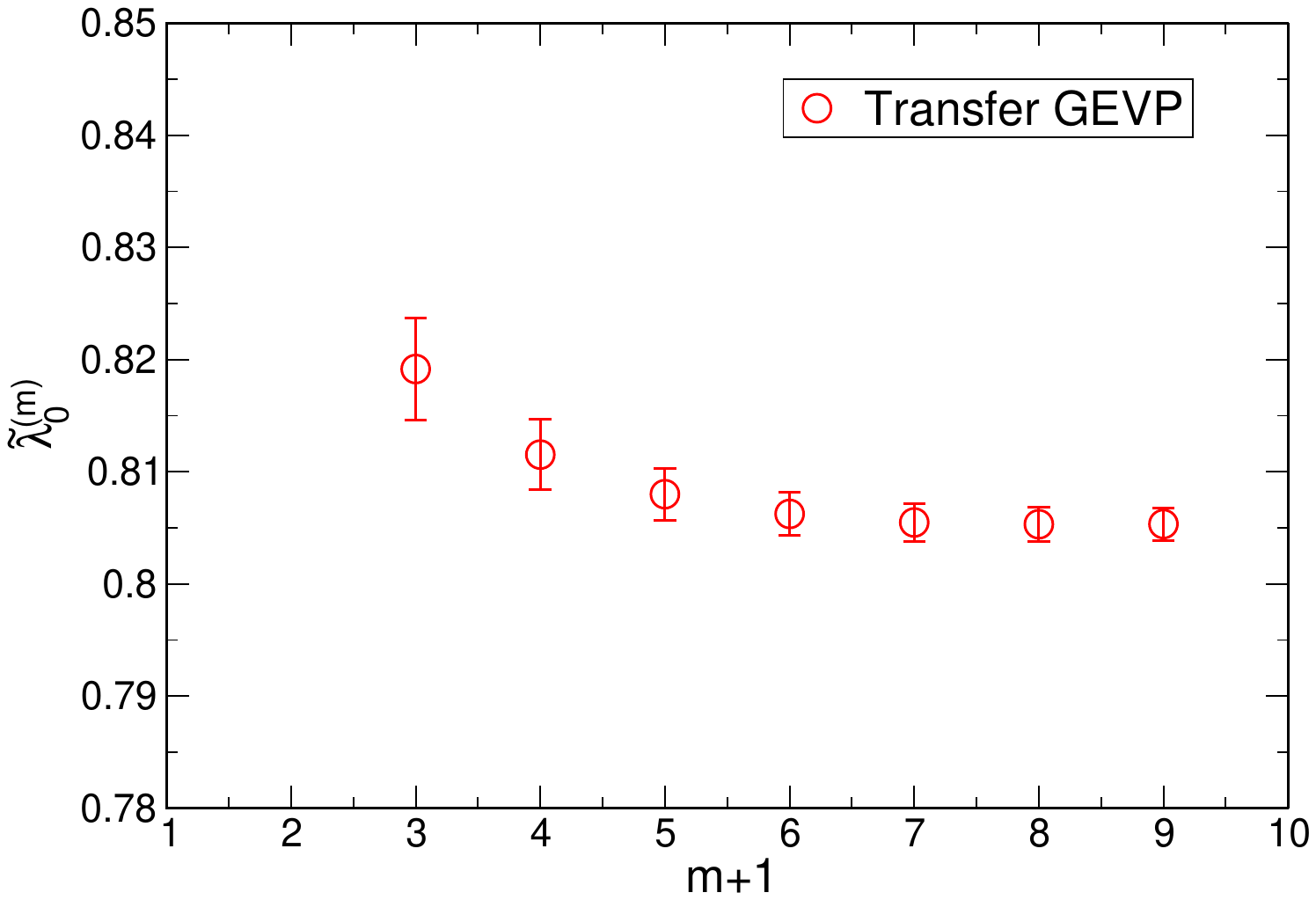}
\caption{
    Estimate of $\tilde{\lambda}_0^{(m)}$ for the $K$-meson correlator. The results are obtained by an extrapolation to $\Delta\lambda^{(m)}_0\to 0$ using $\lambda_0'^{(m)}$ of $r=0$ and 1 for each $m+1$.
}
\label{fig:evLL_s-l_bootstrap}
\end{figure*}

The result of the extrapolation of $\lambda_0'^{(m)}$ to $\Delta\lambda^{(m)}_0\to 0$, denoted as $\tilde{\lambda}_0^{(m)}$, is shown in Figure~\ref{fig:evLL_s-l_bootstrap} for the $K$-meson two-point function. The extrapolation is performed using the data of $r=0$ and 1. It shows a significant dependence on $m+1$ for $m+1\le 5$, suggesting that the size of the Krylov subspace is not sufficient to reproduce the whole Hilbert space. We take the final result from $m=8$ where $\tilde{\lambda}_0^{(m)}$ is already saturated. The result is $\tilde{\lambda}_0^{(8)}=0.8053(14)$, which corresponds to $\tilde{E}_0=-\mathrm{ln}{\tilde{\lambda}_0^{(8)}} = 0.2165(18)$. The distribution of the bootstrap samples is consistent with the normal distribution, and the estimation of the statistical error is robust, {\it i.e.} independent of the methods. The final result for the ground-state energy is compatible with the standard analysis of the correlator using a fit with the exponential function: 0.2143(17).

%
% FIG.
%

%
% FIG.
%

\begin{figure*}[tbp]
\centering
\includegraphics[width=0.7\textwidth,bb=0 10 745 550,clip]{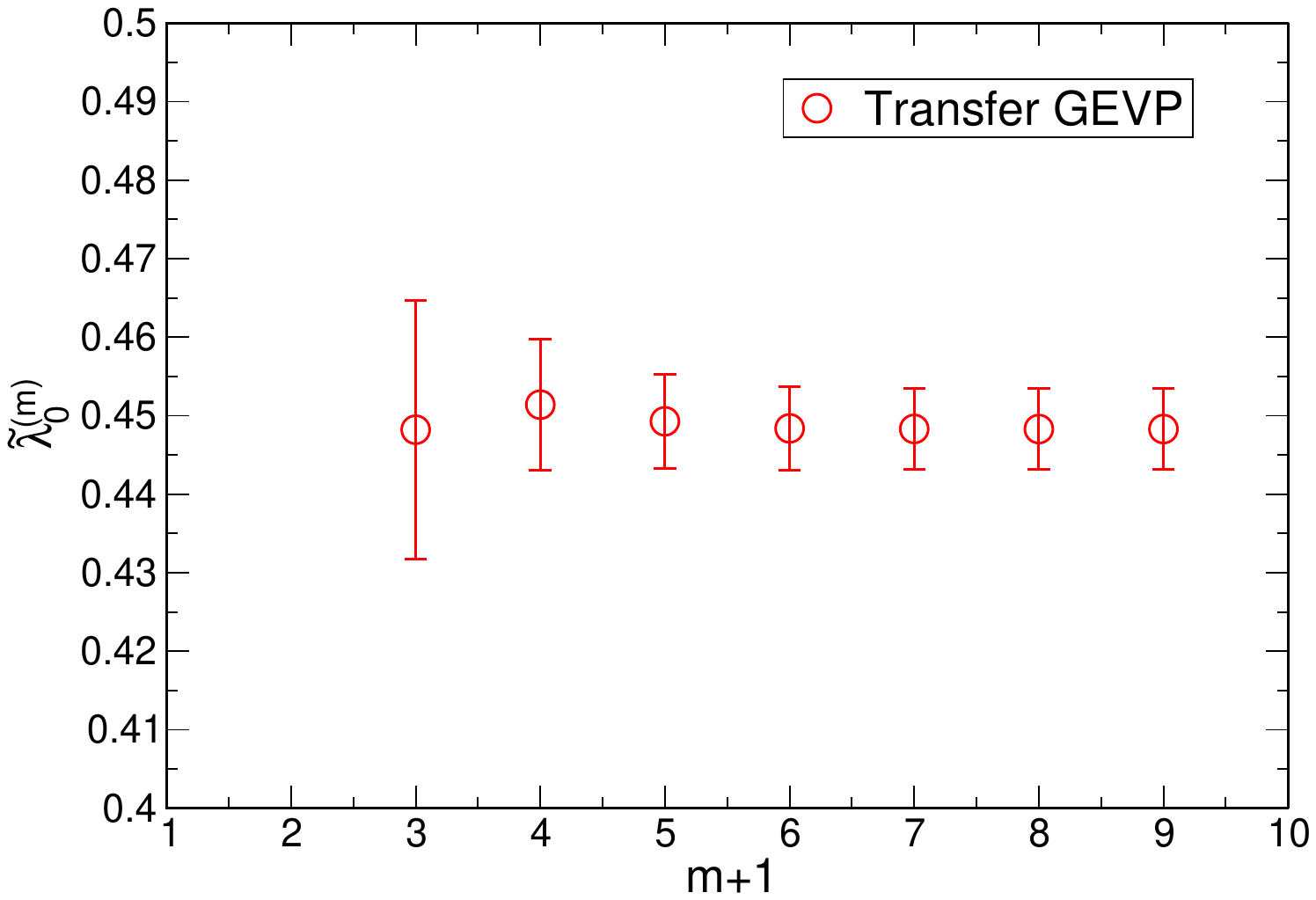}
\caption{
Same as Fig.~\ref{fig:evLL_s-l_bootstrap} for $D_s$.
The extrapolations are performed by using data with $r=1$ and 2.
}
\label{fig:evLL_c-s_bootstrap}
\end{figure*}

We perform the same analysis for the $D_s$-meson correlators. Figure~\ref{fig:evLL_c-s_bootstrap} shows the results of the linear extrapolation using the data at $r=1$ and 2.
The result is $\tilde{\lambda}^{(8)}_0=0.4483(52)$, which corresponds to $\tilde{E}_0=-\mathrm{ln}{\tilde{\lambda}^{(8)}}_0=0.8023(115)$. 
If we use three points, $r$ = 0, 1, and 2, to perform quadratic extrapolation to $\Delta\lambda_0^{(8)}\to 0$, we obtain $\tilde{\lambda}^{(8)}_0=0.4444(48)$, which gives $\tilde{E}_0=-\mathrm{ln}{\tilde{\lambda}^{(8)}}_0=0.8112(107)$.
This result is consistent with the estimate from the standard (multi-)exponential fit: 0.8165(11), albeit the estimated error is significantly larger. Again, this larger statistical error is the cost of not using the assumption of the exponential form.

%
% FIG.
%

\subsection{Ground-state hadron matrix element}
\label{ssec:groud-state_hadron_matrix_element}

In this work, we study the matrix element $\langle D_s|V_4 |D_s\rangle$ for a demonstration of our strategy to construct the ground state. The analysis proceeds as performed for the mock data in Section~\ref{sec:numerical_test_with_mock_data}.

\begin{figure*}[tbp]
\centering
\includegraphics[width=0.7\textwidth,bb=0 30 745 550,clip]{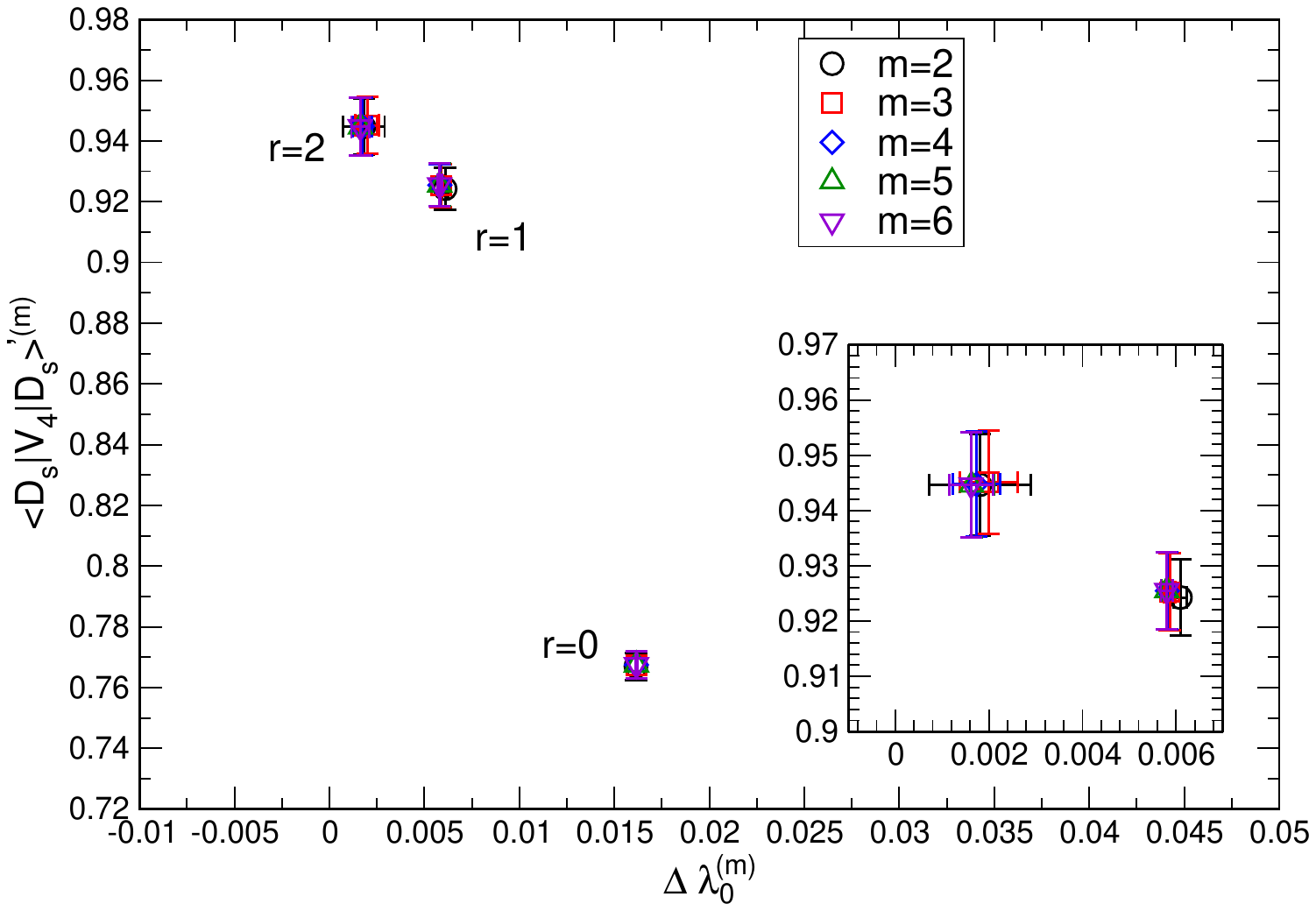}
\caption{
    Matrix element $\langle D_s|V_4 |D_s\rangle$ reconstructed from correlators. It is plotted as a function of $\Delta\lambda_0'^{(m)}$. The data are constructed with $r=0$, 1, 2 for various $m$ shown in the plot.
}
\label{fig:meLL_c-c_GammaTGamma5_DLbootstrap}
\end{figure*}

Figure~\ref{fig:meLL_c-c_GammaTGamma5_DLbootstrap} shows the matrix element $\langle D_s|V_4 |D_s\rangle$ with the ground state reconstructed from correlators themselves. The plot is obtained with $r=0$, 1 and 2 for $m=2,\cdots, 6$. We find a rather strong dependence on $r$, and an extrapolation to $\Delta\lambda_0^{(m)}\to 0$ is important.

%
% FIG.
%

\begin{figure*}[tb]
\centering
\includegraphics[width=0.8\textwidth,bb=0 30 545 450,clip]{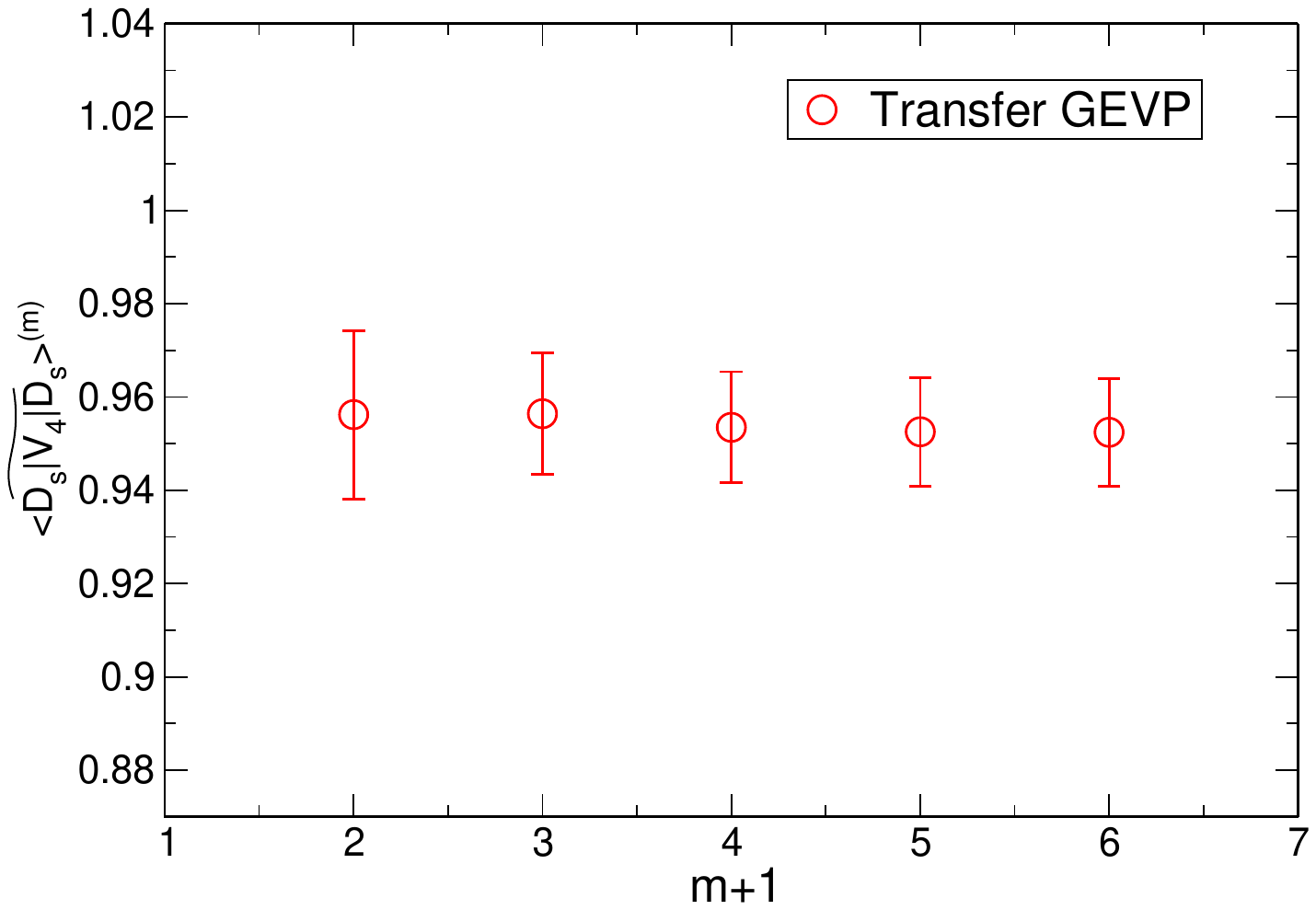}
\caption{
    Matrix element $\widetilde{\langle D_s|V_4 |D_s\rangle}^{(m)}$ after the extrapolation to $\Delta\lambda_0^{(m)}\to 0$. The extrapolations are performed by using data with $r=1,2$.
}
\label{fig:meLL_c-s_GammaTGamma5_bootstrap}
\end{figure*}

In Figure~\ref{fig:meLL_c-s_GammaTGamma5_bootstrap}, the result of the extrapolation is shown. It is obtained from the data at $r=1$ and 2. We find no significant dependence of $\langle D_s|V_4 |D_s\rangle$ on the size of the subspace $m$. The numerical result is $\langle D_s|V_4 |D_s\rangle$ = 0.952(12). If we include the point at $r=0$ to perform a quadratic extrapolation, we obtain 0.946(12), which is consistent with the linear extrapolation with $r=1$ and 2. The statistical distribution is confirmed to be consistent with the normal distribution. Using the standard plateau method, we obtain 0.937(20), and the summation method to suppress the contamination from excited states \cite{Capitani:2012gj} yields 0.954(14). All of these are consistent with each other, and the statistical error is also about the same size.

In continuum theory, $\langle D_s|V_4|D_s\rangle$ is normalized to 1 as a result of current conservation. On the lattice, it is slightly violated due to the  discretization effect, and this size of deviation is reasonable compared to the result for the $B$ meson \cite{Hashimoto:2017wqo}.

%
% FIG.
%

Recall that the matrix element is obtained only from the linear combination of the data without the need of a fitting procedure assuming the ground-state dominance. This indicates that our method provides a useful crosscheck of the standard analysis when the dominance of the ground state is not guaranteed.

%%%%%%%%%%%%%%  SEC 6 %%%%%%%%%%%%%%%%%%%%%%%%
\section{Summary}
\label{sec:summary}

We have presented a new approach to estimate the ground-state mass and matrix element based on a diagonalization of the transfer matrix constructed from the correlation function of the target hadron.
Approaches utilizing the transfer matrix have been proposed in recent years~\cite{Wagman:2024rid, Hackett:2024xnx, Hackett:2024nbe, Chakraborty:2024scw, Ostmeyer:2024qgu, Abbott:2025yhm, Ostmeyer:2025igc}, but discrimination of physical states from spurious modes remained a difficult problem.
In this paper, we propose a new strategy to filter out spurious eigenvalues and estimate the ground-state mass and matrix element.

Our strategy consists of two techniques, the low-rank approximation with the singular-value decomposition and the eigenvalue-variance extrapolation.
The low-rank approximation with the SVD filters out spurious eigenvalues, and potential bias due to the low-rank approximation can be corrected by the eigenvalue-variance extrapolation.
We examine the extrapolation procedure for both the ground-state mass and the matrix element.

The numerical test with the noiseless data explicitly shows that the ground-state mass is reproduced.
This is because the ground-state eigenvalue depends linearly on the eigenvalue-variance $\Delta\lambda_0^{(m)}$, as suggested in \cite{doi:10.1143/JPSJ.69.2723, doi:10.1143/JPSJ.70.2287}.
In addition, we investigated the ground-state matrix element using our strategy and found that the linear extrapolation with respect to $\Delta\lambda_0^{(m)}$ reproduces the matrix element precisely, although the dependence on $\Delta\lambda_0^{(m)}$ could be significantly different depending on the off-diagonal matrix elements.
With the noisy data, we observed that our strategy successfully filters out spurious eigenvalues and reproduces the exact ground-state mass and matrix element within the statistical error estimated by various methods.

A numerical experiment with real lattice data has also been performed. The results of the ground-state mass and matrix element are statistically consistent with the results obtained with the standard plateau method and the summation method.

The virtue of our strategy is that the ground-state saturation can be explicitly tested by the data themselves, while it is assumed in the standard exponential fits of the lattice correlator. This can be a delicate point for more realistic lattice data with larger statistical error when the available subspace is limited. Such studies are left for future work.

%--- acknowledgments ------------------------------------------------  
\begin{acknowledgments}
We thank Mike Wagman for helpful discussions. Numerical calculations in this work were performed on the Supercomputer Fugaku (Project ID: hp240259, hp250224) provided by RIKEN Center for Computational Science (R-CCS).
R.T. is supported in part by JSPS KAKENHI Grant Number JP25KJ0404.
The works of S.H. are supported in part by JSPS KAKENHI Grant Numbers 22H00138 and by the Post-K and Fugaku supercomputer project through the Joint Institute
for Computational Fundamental Science (JICFuS).

\end{acknowledgments}
%----------------------------------- appendix -------------------------------------------------------  
\appendix

\clearpage
\section{Bootstrap resampling to evaluate statistical uncertainty}
\label{app:bootstrap_resampling_to_evaluate_statistical_uncertainty}

Consider measurements of $X(t)$ on $N$ configurations:
\begin{align}
    \{X(t)\}
    =
    \{
    X_1(t),X_2(t),\cdots,X_N(t)
    \}.
\end{align}
We estimate a mean $\hat{\theta}(t)$ of $X(t)$.
In bootstrap resampling, we create a new data set $\{X^*(t)\}$ of the same size $N$
by random sampling with replacement of $\{X(t)\}$. It gives a mean $\hat{\theta}^*(t)$. Repeating this process $N_B$ times, we build a distribution of the means 
\begin{align}
    \{\hat{\theta}^*(t)\}
    =
    \{
    \hat{\theta}_1^*(t),\hat{\theta}_2^*(t),\cdots,\hat{\theta}_{N_B}^*(t)
    \}.
\end{align}
The mean and error are then estimated as
\begin{align}
    \langle \hat{\theta}^{*}(t) \rangle
    & =
    \frac{1}{N_B}
    \sum_{i=1}^{N_B}
    \hat{\theta}_i^*(t), \\
    \delta \hat{\theta}^*(t)
    & =
    \sqrt{ \{ \langle (\hat{\theta}^*(t))^2 \rangle - \langle \hat{\theta}^*(t) \rangle^2 \} }.
\end{align}
Note that this error estimation is made under an assumption that the means have a normal distribution. It is valid in the limit of large $N$, but for finite $N$ the confidence interval (CI) may provide more robust estimate. 

The most common CI is the percentile interval 
\begin{align}
    \label{eq:bootstrap_ci}
    \mathrm{CI_{bootstrap}} = \left[\hat{\theta}^*_{(\alpha/2)}(t), \hat{\theta}^*_{(1-\alpha/2)}(t)  \right]
\end{align}
with the significance level $\alpha$, which defines the probability that the true value lies outside the CI. Practically, $\alpha=0.32$ corresponds to 68\% CI, which is identical to one standard deviation, provided that the distribution of the bootstrap samples is well described by the normal distribution. The bootstrap CI does not assume the distribution and thus provides a robust way to estimate the uncertainty.

To approximate quantiles of non-normal distribution, the Cornish-Fisher (CF) expansion is often used. It utilizes the information of cumulants (skewness $\kappa_3$, kurtosis $\kappa_4$, etc.) of the distribution to adjust the quantiles. 
The adjusted quantile $q_\alpha$ is given by
\begin{align}
    \label{eq:CF_quantile}
    q_\alpha \nonumber
    \simeq z_\alpha 
    + \frac{1}{6}(z_\alpha^2-1)\kappa_3
    + \frac{1}{24}(z_\alpha^3-3z_\alpha)\kappa_4
    - \frac{1}{36}(2z_\alpha^3-5z_\alpha)\kappa_3^2
\end{align}
with $\alpha$-quantile of the standard normal distribution $z_\alpha$.
With the adjusted quantile $q_\alpha$, the CI is constructed as
\begin{align}
    \mathrm{CI_{CF}} = \langle \hat{\theta}^*(t) \rangle \pm \delta \hat{\theta}^*(t)\times q_\alpha,
\end{align}
where $\alpha=0.32$ corresponds to the 68\% CI.
Recall that the CF approach assumes that the underlying distribution is unimodal and close to normal. Hence, it has limitations when the distribution is highly skewed or heavy-tailed.

\section{Reconstructed correlation function}
\label{app:reconstructed_correlation_function}
Using the eigenvalues and eigenstate vectors of the GEVP, one can also reconstruct the two-point correlation functions. The full correlation function can be represented as a sum of the contributions from the ground state and excited states (spectral decomposition):
\begin{align}
    \label{eq:twopt_spectraldecomposition}
    C(t)
    & =
    \langle \psi |  
    \hat{\mathcal{T}}^{t}
    | \chi \rangle
    \nonumber
    \\
    & =
    \sum_{n,n'}
    \langle \psi |E_n\rangle\langle E_{n'} |\chi\rangle
    \langle E_n  |\hat{\mathcal{T}}^{t} | E_{n'} \rangle
    \nonumber
    \\
    & =
    \sum_n \langle \psi | E_n \rangle  \langle E_n|\chi\rangle (\lambda_n)^t
\end{align}
with ${1}=\sum_n|E_n\rangle \langle E_n |$, $\langle E_n | E_{n'} \rangle = \delta_{nn'}$, and $\hat{\mathcal{T}}|E_n\rangle = \mathrm{e}^{-E_n}|E_n\rangle$.
Therefore, the correlation function corresponding to a specific state $n$ can be reconstructed as
\begin{align}
    C_n^{(m)}(t)
    & = 
    \langle \psi | E_n \rangle  \langle E_n|\chi\rangle 
    (\lambda_n)^t
    \nonumber
    \\
    \label{eq:reconst_correlation_function}
    & =
    \frac{
    \sum_{j=0}^m
    \left( y^{(m),L}_n \right)_j C(j)}{\sqrt{\langle x^{(m),L}_n|x^{(m),R}_n\rangle}}
    \frac{
    \sum_{j=0}^m
    \left( y^{(m),R}_n \right)_j C(j)}{\sqrt{\langle x^{(m),L}_n|x^{(m),R}_n\rangle}}
    (\lambda_n)^t.
\end{align}
We note that no fitting is necessary to reconstruct the correlation function in this approach. Excited states can also be reconstructed in the same way, insofar as the GEVP can be performed with sufficient accuracy.

%----------------------------------- bibliography -------------------------------------------------------  
\bibliography{TransferMatrixGEVP/main}
\bibliographystyle{man-apsrev}
 
\end{document}